\documentclass[12pt]{iopart}

\usepackage{iopams}
\usepackage{graphicx}
\usepackage{hyperref}

\usepackage{amsfonts}
\usepackage{amssymb}
\usepackage{mathrsfs}  

\usepackage[colorinlistoftodos]{todonotes}
\usepackage{epstopdf}

\def\be{\begin{equation}}
\def\ee{\end{equation}}
\def\ba{\begin{eqnarray}}
\def\ea{\end{eqnarray}}

\def\go{\mathring{q}}
\def\epsilono{\mathring{\epsilon}}

\def\SU(2){\rm SU(2)}
\def\su(2){\rm su(2)}

\def\h{\hat}

\def\rmd{\mathrm{d}}
\def\f{\frac}

\def\h{\hat}
\def\lp{\ell_{\rm Pl}}
\def\rhop{\rho_{\rm Pl}}
\def\rcr{\rho_{\rm sup}}

\def\C{\mathcal{C}}
\def\D{\mathcal{D}}
\def\F{\mathcal{F}}
\def\G{\mathcal{G}}
\def\H{\mathcal{H}}

\def\man{\mathcal{M}}
\def\S{\mathcal{S}}
\def\V{\mathcal{V}}
\def\P{\mathcal{P}}
\def\SA{{\it SA }}

\def\gps{\mathbf{\Gamma}}
\def\Al{\mathfrak{A}}
\def\Alred{\mathfrak{A}_{\rm Red}}

\def\Hkg{{\mathcal{H}}^{\rm kin}_{\rm grav}}
\def\Hpg{{\mathcal{H}}^{\rm phys}_{\rm grav}}
\def\Hkr{{\mathcal{H}}^{\rm kin}_{\rm Red}}

\newcommand{\pb}[1]{\hbox{\lower0.5ex\hbox{${}_{\leftarrow}$}}\kern-1.9ex{#1}}

\begin{document}

\title{A Short Review of Loop Quantum Gravity}

\author{Abhay Ashtekar}
\address{Institute for Gravitation \& the Cosmos, and
    Physics Department, Penn State, University Park, PA 16802,
    USA}
\ead{ashtekar.gravity@gmail.com}
\author{Eugenio Bianchi}
\address{Institute for Gravitation \& the Cosmos, and
    Physics Department, Penn State, University Park, PA 16802,
    USA}
\ead{ebianchi@psu.edu}

\begin{abstract}
An outstanding open issue in our quest for physics beyond Einstein is the unification of general relativity (GR) and quantum physics.  Loop quantum gravity (LQG) is a leading approach toward this goal. At its heart is the central lesson of GR: Gravity is a manifestation of spacetime geometry. Thus, the approach emphasizes the quantum nature of geometry and focuses on its implications in extreme regimes -- near the big bang and inside black holes-- where Einstein's smooth continuum breaks down. We present a brief overview of the main ideas underlying LQG  and highlight a few recent advances. This report is addressed to non-experts.
\end{abstract}

\section{Introduction}
\label{s1}

Einstein emphasized the necessity of a quantum extension of general relativity (GR) already in his 1916  paper \cite{einstein1916}  on gravitational waves, where he said {\it ``\ldots   it appears that quantum theory would have to modify not only Maxwellian  electrodynamics, but also the new theory of gravitation.''}  A century has passed since then, but the challenge still remains. So  it is natural to ask why the task is so difficult. Generally the answer is taken to be the lack of experimental data with direct bearing on quantum aspects of gravitation. This is certainly a major obstacle. But this cannot be the entire story.  If it were, the lack of observational constraints  should have led to a plethora of theories and the problem should have been that of narrowing down the choices. But the situation is just the opposite: As of now we do not have a single satisfactory candidate! 

The central reason, in our view, is quite different. In GR,  gravity is encoded in the very geometry of  spacetime. And its most spectacular predictions  --the big bang, black holes and  gravitational waves--  emerge from this encoding. Indeed, to create GR, Einstein had to begin by introducing a \emph{new syntax} to describe all of classical physics: that of Riemannian geometry. Thus, spacetime is represented as a 4-dimensional manifold $\man$ equipped with a (pseudo-)Riemannian metric $g_{ab}$ and matter is represented by tensor fields.  To construct a quantum theory of gravity, we need yet another, newer syntax --that of a \emph{quantum} Riemannian geometry-- where one only has a probability amplitude for various spacetime geometries in place of a single metric.  Creation of this syntax was truly challenging because all of twentieth physics presupposes a classical spacetime with a metric, its sharp light cones, precise geodesics and proper time assigned to clocks. How do we do physics if we do not have a specific spacetime continuum in the background to anchor the habitual notions we use? A basic premise of  loop quantum gravity (LQG) is that it is these  conceptual issues that have posed the main obstacle in arriving at a satisfactory quantum gravity theory. 

Therefore, the first step in LQG was to systematically construct a specific theory of quantum Riemannian geometry, a task completed in 1990s  (see, e.g., \cite{alrev}-\cite{30years:KG}). 
The new syntax  arose from two principal ideas: (i) A reformulation of GR  (with matter) in the language of gauge theories  --that successfully describe the other three basis forces of Nature--  but now \emph{without reference to any background field}, not even a spacetime metric; and, (ii) Subsequent passage to quantum theory using non-perturbative techniques from gauge theories --such as Wilson loops--  again without reference to a background.  Now, if  a theory has no background field,  it has access only to an underlying manifold and must therefore be \emph{covariant with respect to diffeomorphisms} --the transformations that preserve the manifold structure.  As we will explain, diffeomorphism covariance together with non-perturbative methods naturally lead to a fundamental, in-built discreteness in geometry that foreshadows ultraviolet finiteness. Continuum arises only as a  coarse grained approximation. The familiar spacetime continuum \`a la  Einstein is \emph{emergent} in two senses. First, it is built out of certain fields that feature naturally in gauge theories, without any reference to  a spacetime metric. Second,  it emerges only \emph{on coarse graining} of the fundamental discrete structures --the `atoms of geometry'-- of the quantum Riemannian framework.

Over the past two decades, this new syntax has been used to address some of the key conceptual issues in quantum gravity that have been with us for over half a century \cite{dirac2,wheeler}.  To begin with, matter and geometry are both quantum mechanical `at birth'.  A matter field $\hat\phi$ propagates on a quantum geometry represented by a wave function $\Psi$ of geometries.  If $\Psi$ is sharply peaked, to leading order the dynamics of $\hat\phi$ is well approximated by that on a classical spacetime metric at which it is peaked. But to the next order, it is also sensitive to the quantum fluctuations of geometry. Since there is no longer a specific metric $g_{ab}$,  we do not have a sharp notion of, e.g., `proper time'.   Quantum dynamics is relational: Certain degrees of freedom --such as a matter field, for example-- can be used as a relational clock with respect to which other degrees evolve.  In the gravitational sector, the ultraviolet regularity is made manifest by taming of the most prominent singularities of GR. In cosmology,  the big bang and big crunch singularities are replaced by a big bounce.  (In fact, all strong curvature singularities are tamed  \cite{ps1}, including the `big-rip'-type 
and `sudden death'-type that cosmologists often consider (see. e.g., \cite{30years:IAPS}, or, \cite{asrev}).)
As a result, in loop quantum cosmology (LQC), spacetimes do not end at the big bang or the big crunch. Rather, quantum geometry extends the spacetime to other macroscopic branches.  Similarly, because the quantum spacetime can be much larger than what GR  would have us believe, there is now a new avenue for `information recovery' in the black hole evaporation process. 

There are also other approaches to quantum gravity, each emphasizing certain issues and hoping that the remaining ones, although important, will be addressed rather easily once the `core' difficulties are resolved (see, e.g., Chapters 11 and 12 in \cite{centennial}). At first  the divergence of subsequent developments seems surprising. However, as C. N. Yang \cite{yang} has explained: {\it ``That taste and style have so much to do with physics may sound strange at first, since physics is supposed to deal objectively with the physical universe. But the physical universe has structure, and one's perception of this structure, one's partiality to some of its characteristics and aversion to others, are precisely the elements that make up one's taste. Thus it is not surprising that taste and style are so important in scientific research.''}  For example,  because string theory  was developed by particle physicists, the initial emphasis was on \emph{unification} of all interactions, including gravity.  To achieve this central goal,  radical departure from firmly established physics was considered a small price to pay. Thus, higher spacetime dimensions, supersymmetry and a negative cosmological constant were introduced as \emph{fundamental ingredients} in the 1980s and 1990s, with a hope that evidence for these extrapolations would be forthcoming.  So far, these hopes have not been realized, nor has the initial idea of unification been successful from a phenomenological perspective \cite{ias}. However, the technical simplifications brought about by these assumptions \emph{have} led to unforeseen mathematical results, facilitating explorations in a number of areas that are not directly related to quantum gravity. Another significant development is the `asymptotic safety program' that is now providing some insights into potential quantum gravity implications on the standard model of particle physics \cite{eichhorn}.  Loop quantum gravity, by contrast, has focused on the fundamental issues of quantum gravity proper that have been with us since the initial investigations by Bergmann \cite{bergmann}, Dirac \cite{dirac1,dirac2}, Wheeler \cite{wheeler} and others: How do  we `quantize' constrained Hamiltonian systems without introducing background fields or perturbative techniques? What does dynamics mean if there is no spacetime metric in the background? Can we successfully calculate `quantum transition amplitudes'?  Since diffeomorphisms can move just  one of given $n$ points keeping the remaining $n-1$ fixed, can there be non-trivial $n$-point functions in a diffeomorphism covariant theory?  Are the curvature singularities of classical GR  naturally resolved by quantum gravity? Are the ultraviolet divergences of quantum field theory (QFT) cured?  A large number of researchers have addressed these issues over the last two decades. Since there are literally thousands of papers on the subject, we will not even attempt to present a comprehensive bird's eye view.  Rather, following the goal of the ``Key Issue Reviews'' of the journal,  this article is addressed to non-experts and provides a broad-brush portrait of the basic underlying ideas and illustrates the current status through a few examples. 

This discussion is organized as follows. Section 2 introduces quantum geometry and Section 3 discusses the current status of quantum dynamics. Section 4 presents an illustrative application: to the cosmology of the early universe. We conclude in Section 5 with a brief discussion of some of the advances that could not be included, limitations of the current status and the key open issues.

\section{Quantum Riemannian Geometry}
\label{s2}
\bigskip\bigskip

As explained in Section \ref{s1}, two key ideas led to a detailed quantum theory of geometry. We present them in the two subsections that follow and discuss salient features of the fundamental discreteness of geometry that emerges.

\subsection{Gauge theory notions simplify GR }
\label{s2.1}

Recall that a solution to the equations of motion of a point particle can be visualized as a trajectory in the configuration space of its positions. In GR, spatial metrics $q_{ab}$ (of signature $+,+,+$) on a 3-dimensional manifold $M$ represent configurations of spacetime geometries, and a solution to Einstein's equation can be viewed as a trajectory in the (infinite dimensional) configuration space $\C$ of all $q_{ab}$'s. Following Wheeler,  $\C$ is called  \emph{superspace.}  (Note that this is unrelated to the superspace introduced in supergravity.) The conjugate momenta are tensor fields $p_{ab}$ (with density weight one).  Following the lead of Bergmann and Dirac,  Arnowitt-Deser-Misner (ADM) introduced a Hamiltonian description of GR  (for a summary, see \cite{adm}).  Einstein's equations break up into 4 constraint equations on the pair  $(q_{ab}, p_{ab})$ --that involve no time derivatives--  and six evolution equations that dictate how this canonically conjugate pair evolves, providing us with dynamics. This framework came to be called \emph{geometrodynamics}. 

For simplicity, let us consider source-free Einstein equations and assume that the 3-manifold $M$ is compact since removal of these restrictions only makes the equations more complicated without changing the essential points in our discussion. The constraint equations naturally split into a (co-)vectorial part $C_{a}$ and a scalar part $C$ on $M$:
\be \label{admcon} \hskip-1.5cm
C_{a} := -2 q_{ac} D_{b}\, p^{ac} =0, \quad {\rm and} \quad   C := - {q}^{\f{1}{2}} \,\mathcal{R} - \epsilon\, q^{-\f{1}{2}} \big(q_{ac} q_{bd} - \,{\textstyle{\f{1}{2}}} q_{ab} q_{cd}\big)\,  p^{ab}\, p^{cd}\,  =\, 0, \,\, \ee
where $D$ is the covariant derivative operator of the metric $q_{ab}$;  $q$, its determinant; and $\mathcal{R}$, its scalar curvature;  and $\epsilon=1$  in the  Riemannian signature $+,+,+,+$ and $-1$ in the Lorentzian signature $-,+,+,+$. Note that because $C^{a}$ and $C$ are \emph{fields} on $M$ we have an infinite number of constraints; 4 per points of $M$.  The Poisson bracket between any two of them vanishes on the constraint surface in the phase space; so the constraints are said to be of \emph{first class} in Dirac's terminology.  Since the configuration variable $q_{ab}$ has six components at each point of $M$, we have $6-4 = 2$ true degrees of freedom of the gravitational field in GR .

It turns out that the  canonical transformations generated by $C^{a}$ correspond to spatial diffeomorphisms on $M$, whence it is called the \emph{Diffeomorphism constraint}. Simialrly, those generated by $C$ correspond to time evolution (in the direction normal to $M$,  when it is embedded in spacetime $(\man, g_{ab})$). Hence $C$ is called the \emph{Hamiltonian constraint}. Thus,  the Hamiltonian generating evolution in a generic time-like direction is a linear combination of constraints, 
\be \label{admH}  \bar{H}_{{N}, {\vec{N}}} (q,p) := \int_{M}  \big({N}C\, +\, {N}^{a}\,C_{a} \, \big)\, \rmd^{3}x\ee 
\noindent where the freely specifiable positive function $N$  on the 3-manifold $M$ is called the \emph{lapse} and the freely specifiable vector field $N^{a}$, the \emph{shift}. ($H_{N, \vec{N}}$ is independent of the choice of coordinates on $M$ because the integrand is a density of weight 1.)  The form (\ref{admH}) of the Hamiltonian just reflects the fact that GR  is a background independent --or fully covariant-- theory. Different choices of $N,  N^{a}$  yield the same solution, presented with different constant-time slices and of the vector field defining time evolution. Note that, because of the presence of $D, q, q^{ab}$, and $\mathcal{R}$, the constraints $C$ and $C^{b}$ are rather complicated, non-polynomial functions of the basic canonical variables $q_{ab}, p^{ab}$. As a result the equations of motion they generate are also quite complicated: Setting $N^{a}=0$ for simplicity,  equations governing `pure' time evolution are
\ba \label{admevo}
\dot{q}_{ab} &=& 2 N q^{-\f{1}{2}} \big(q_{ac} q_{bd} -\f{1}{2}  q_{ab} q_{cd})\, p^{cd}\, ,
\nonumber\\ 
\dot{p}^{ab} &=&  \epsilon\, q^{\f{1}{2}}\big(q^{ac} q^{bd} - q^{ab} q^{cd} \big) D_{c}D_{d} N - \epsilon\, q^{\f{1}{2}} N \big(q^{ac} q^{bd} - \f{1}{2} q^{ab} q^{cd} \big) \,\mathcal{R}_{cd} \nonumber\\
&-&  q^{-\f{1}{2}}  N\,\big(2\delta^{a}_{d} \delta^{b}_{n}q_{cm} - \delta^{a}_{m} \delta^{b}_{n}q_{cd} - \f{1}{2} q^{ab} (q_{cm} q_{dn}  - \f{1}{2} q_{cd} q_{mn}) \big) p^{cd} p^{mn}. 
 \ea
 Details are unimportant but the complexity of these equations is evident. It has been the major reason why equations of \emph{quantum} geometrodynamics have yet to be given a mathematically precise meaning; they continue to remain formal even today.

Let us now change gears and turn to gauge theories. We will introduce a \emph{background independent} Hamiltonian framework ab initio using general considerations and relate it to geometrodynamics at the end \cite{aamv}. In gauge theories, the configuration variable is a connection --or a vector potential-- $A_{a}^{i}$ on $M$, where the index $i$ refers to the Lie algebra of the gauge group, which we will take to be $\SU(2)$.  The conjugate momenta are electric fields $E^{a}_{i}$ which are vector fields on $M$ (with density weight $1$) that also take values in the Lie algebra $\su(2)$ of $\SU(2)$. The Cartan-Killing metric $\go_{ij}$  enables one to freely raise and lower the internal indices $i,j,k \ldots$,  and we can also use the structure constants $\epsilono^{i}{}_{jk}$ to construct the curvature/field strength, $F_{ab}^{i} := 2 \partial_{[a}A_{b]}^{i} + \epsilono^{i}{}_{jk} A_{a}^{j} A_{b}^{k}$, which is gauge covariant.  So, the phase space $\gps$ is the same as in the theory of weak interactions. However, we no longer have a spacetime metric in the background. Therefore the symmetry group of the theory will be generated by the local $\SU(2)$ gauge transformations that leave each point of $M$ invariant, \emph{and the diffeomorphisms}, motions on $M$ that respect just its differential structure. Therefore, we are led to ask for the simplest gauge covariant functions of  the canonical variables $(A_{a}^{i}, \, E^{a}_{i})$ that do not contain any background fields (not even a  metric). The simplest ones --that contain $A_{a}^{i}$ and $E^{a}_{i}$ at most quadratically are:
\be \label{con1} 
\G_{i} := \D_{a} E^{a}_{i} ; \quad  \V_{a} := E^{b}_{i} F_{ab}^{i};\quad {\rm and} \quad \S :=  {\textstyle{\f{1}{2}}}\, \epsilono^{ij}{}_{k}\, E^{a}_{i} E^{b}_{j} F_{ab}^{k} \, ; \ee
where $\D_{a}E^{a}_{i} := \partial_{a}E^{a}_{i} +\epsilono_{ij}{}^{k}    A_{a}^{j} E^{a}_{k}$. 
Note that the Gauss constraint of the gauge theory that generates the local $\SU(2)$ rotations is precisely $\G_{i}=0$. This feature and the fact that that $\V_{a}$ is a (co-)vector field on $M$ and $S$ a scalar field --similar in structure to $C_{a}, C$ of geometrodynamics-- motivate the introduction of seven constraints on the gauge theory phase space $\gps$:
\be \label{con2} 
\G_{i} =0;\qquad \V_{a} =0; \qquad  \S =0. \ee
One can again check that these are of first class in Dirac's terminology \cite{aa-newvar, aa-book}.  
Since now the configuration variable is $A_{a}^{i}$ with nine components, and we have seven first class constraints (\ref{con2}), there are again two true degrees of freedom. Since the theory is background independent, let us introduce a Hamiltonian $H(A,E)$ as a linear combination of the constraints (\ref{con2}):
\be \label{H} H_{N, \vec{N}, N^{i}}\,(A,E) := \int_{M} \big(N \S + N^{a}\V_{a} + N^{i} \G_{i}\,  ) \rmd^{3}x \ee
where the freely specifiable scalar $N^{i}$ with an $\su(2)$ index $i$ is a generator of gauge rotations, $N^{a}$ is again the shift  and $N$, the lapse. (However, because $S$ is scalar with density weight $2$, the lapse $N$ is a scalar with density weight $-1$, rather than a function as in (\ref{admH}).)  Since the Hamiltonian is a low order polynomial in the canonical variables, in striking contrast to (\ref{admevo}),  the evolution equations only involve \emph{low order polynomials}.  Again, setting $N^{a} =0$ and $N^{i} =0$ for simplicity, we obtain:
\be \label{evo} 
{\dot{A} }^i_a=    N \,E^b_j\, F_{ab}^{k}\, \epsilono^{ij}{}_{k}, \qquad {\rm and} \qquad 
{\dot{E} }^a_i=   \D_a \big(N\,E^a_j E^b_k \big) \, \epsilono_{i}{}^{jk}.  \ee
To summarize, although we started with the kinematics of an $\SU(2)$ gauge theory and just wrote down the simplest constraints compatible with $\SU(2)$ gauge invariance  and background independence,  we have again arrived at a \emph{diffeomorphism covariant theory with two degrees of freedom}! So, it is tempting to conjecture that this theory may be related to GR,  where the Riemannian structures are no longer at the forefront as in geometrodynamics, but \emph{emerge} from the background independent gauge theory.  It turns out that the two theories are in fact \emph{equivalent} in a precise sense.  In particular, the unruly evolution equations (\ref{admevo}) of geometrodynamcs are equivalent to the much simpler equations (\ref{evo}). Furthermore, it turns out that the right sides of (\ref{evo}) have a simple geometrical meaning  in the gauge theory framework \cite{aamv}. But this simplicity is lost when the background independent gauge theory is recast using Riemannian geometry.

The salient features of the dictionary for transition to geometrodynamics can be summarized as follows. Let us first consider Riemannian GR  with signature $+,+,+,+$.  Each electric field $E^{a}_{i}$ provides a map from fields $\lambda^{i}$, taking values in  $\su(2)$, to vector fields (with density weight $1$) $\lambda^{a}$ on $M$:\, $ \lambda^{i} \to \lambda^{a} := E^{a}_{i} \lambda^{i}$. Let us restrict ourselves to the generic case when the map is $1$-$1$. Then  each $E^{a}_{i}$ defines a $+,+,+$ metric $q_{ab}$ on $M$ via:\, $q\, q^{ab} = \go^{ij} E^{a}_{i} E^{b}_{j}$ where $q$ is the determinant of $q_{ab}$. Thus, the electric field $E^{a}_{i}$ serves as an orthonormal triad (with density weight one) for the metric $q_{ab}$. What about  the connection $A_{a}^{i}$?  Set $A_{a\,A}{}^{B}  := A_{a}^{i} \tau_{i\, A}{}^{B}$, where  $\tau_{i\, A}{}^{B}$ are Pauli matrices.  Then, the gravitational meaning of $A_{a\,A}{}^{B}$ is the following: It enables us to parallel transport $\SU(2)$ spinors --the left-handed spin 1/2 particles of the standard model-- in the gravitational field represented by the geometrodynamical pair $(q_{ab}, p^{ab})$.  With this correspondence,  $(q_{ab}, \,p^{ab})$ satisfy the geometrodynamical constraints (\ref{admcon}) and evolution equations (\ref{admevo}) if $(A_{a}^{i}, E^{a}_{i})$ satisfy the constraints (\ref{con2}) and the evolution equations (\ref{evo}).  The curvature $F_{ab\, A}{}^{B} := F_{ab}{}^{i}\, \tau_{i\,A}{}^{B}$ in gauge theory has a simple geometrical interpretation: it is the restriction to $M$ of the \emph{self-dual part} of the curvature of the 4-metric representing the dynamical trajectory passing through $(q_{ab}, p^{ab})$.  Furthermore, (\ref{con2}) and  (\ref{evo}) provide a slight generalization of Einstein's equations because they continue to be valid even when $E^{a}_{i}$ fails to be $1$-$1$, i.e., $q_{ab}$ becomes degenerate. While all equations on the gauge theory side are low order polynomials in basic variables, those on the geometrodynamics side have a  complicated non-polynomial dependence simply because $(q_{ab}, p^{ab})$ are complicated non-polynomial functions of  $(A_{a}^{i}, E^{a}_{i})$. Given that electroweak and strong interactions are described by gauge theories, it is interesting that equations of GR  simplify considerably when the theory is recast as a background independent gauge theory, by regarding Riemannian geometry as `emergent'. 

For Lorentzian GR,  we have $q\, q^{ab} = - \go^{ij} E^{a}_{i} E^{b}_{j}$, and the self-dual part of the gravitational curvature and the connection $A_{a}^{i}$ that parallel transports left handed spinors are now complex-valued. To ensure that we recover real, Lorentzian GR  therefore, one has to require that $q_{ab}$ is positive definite and its time derivative is real. (Note from Eq. (\ref{evo}) that $\dot{E}^{a}_{i}$ involves $A_{a}^{i}$.) If the condition is imposed initially, it is preserved in time. (For details, see \cite{aamv}.)  There is an added bonus: the full set of equations (\ref{con2}) and (\ref{evo}) automatically constitutes a symmetric hyperbolic system, making it directly useful to evolve arbitrary initial data using numerical or analytical approximation schemes as well \cite{reula,shinkai}. 

To summarize, one can recover GR  from a natural background independent gauge theory, which has the further advantage that it simplifies the constraints as well as evolution equations enormously.  Also, since the Hamiltonian constraint $\S = \epsilono^{ij}{}_{k}\, E^{a}_{i} E^{b}_{j} F_{ab}^{k}$ is purely quadratic in momenta --without a potential term, solutions to evolution equations have a natural geometrical interpretation as geodesics of the `supermetric' $\epsilono^{ij}{}_{k}\, F_{ab}^{k}$ on the (infinite dimensional)  space of connections.  Results presented in this section have been extended to include the fields  --scalar, Dirac and Yang-Mills-- that feature in the standard model \cite{art,aa-book}. From a gauge theory perspective, then, the Riemannian geometry that underlies GR  can be thought of as a secondary, emergent structure.

\subsection{Background independence implies discreteness}
\label{s2.2}
Considerations of section \ref{s2.1} suggest that the passage to quantum theory would be easier if we use the gauge theory version of GR. Indeed, this version made available several key tools that had not been used in quantum gravity before. Specifically we have the notions of : (i) Wilson lines --or holonomies-- $h_{\ell}(A)$ of the connection $A_{a}^{i}$ that parallel transport a left handed spinor along  curves/links $\ell$; 
\footnote{In the early days of LQG, emphasis was on Wilson \emph{loops} rather than Wilson lines; this is the origin of the name \emph{Loop} Quantum Gravity \cite{loops1,loops2}.}
and, (ii) electric field fluxes $E_{f,S}$, smeared with test fields $f^{i}$, across a 2-surface $S$:
\be h_{\ell}(A) := \P \, \exp \int_{\ell} A_{a}^{i} \tau^{i}  \rmd{\ell}^{a}\qquad {\rm and} \qquad E_{f,S} = \int_{S} f^{i} E^{a}_{i} \rmd^{2} S_{a}\, , 
\label{hAE}
\ee
both defined without reference to a background metric or length/area element.  It turns out that the set of these functions is large enough to provide a (over-complete) coordinate system on the phase space $\gps$ and is also closed under Poisson brackets. Therefore it serves as a point of departure for quantization. Thus we can introduce abstract operators $\h{h}_{\ell}, \, \h{E}_{f,S}$, and consider the algebra $\Al$ they generate. This is the analog of the familiar Heisenberg algebra in quantum mechanics.  The task is to choose a representation of $\Al$. The Hilbert space $\Hkg$ that carries the representation would then be the space of kinematical quantum states  --the quantum analog of the gravitational phase $\gps$ of GR--  serving as the arena to formulate dynamics.

Now, in Riemannian GR,  the $h_{\ell}$ take values in $\SU(2)$ which is compact, while in Lorentzian GR $h_{\ell}$ takes values in $\mathbb{C}\SU(2)$, which is non-compact.  As explained in section \ref{s2.1}, this feature does not introduce any difficulties in the classical theory. However, non-compactness creates obstacles in the rigorous functional analysis that is needed to introduce $\Hkg$. Two strategies have been pursued to bypass this difficulty. In the first, one can begin with Riemannian signature, construct the full theory and then pass to the Lorentzian signature through a quantum version of the generalized Wick transform that maps self-dual connections in the Riemannian section to those in the Lorentzian \cite{wick1}-\cite{wick3}. In the second, and much more widely followed strategy, one makes a canonical transformation by replacing the `$\mathrm{i}$' that features in the expression of the self-dual connection in the Lorentzian theory by a real parameter $\gamma$. Thus, one works with a real connection also in the Lorentzian theory \cite{fb} which, however, is no longer self-dual. $\gamma\,$ is referred to as the \emph{Barbero-Immirzi} parameter of LQG. It represents a $1$-parameter quantization ambiguity, analogous to the $\theta$-ambiguity in QCD \cite{theta1,theta2}. The value of this parameter has to be fixed by observations or thought experiments. Mathematical structures underlying kinematics are the same in both strategies; they differ in their approach to dynamics, discussed in sections \ref{s3} and \ref{s5}.

Let us return, then, to kinematical considerations. In quantum mechanics, the von-Neumann's theorem guarantees that the Heisenberg algebra admits a unique representation satisfying certain regularity conditions (see, e.g., \cite{vonN,hall}). However, in Minkowskian QFTs, because of the infinite number of degrees of freedom, this is not the case in general: The standard result on the uniqueness of the Fock vacuum assumes \emph{free field dynamics} \cite{nonunique,segal}. What is the situation with the algebra $\Al$? Now, in addition to the standard regularity condition, we can and \emph{have to} impose the stringent requirement of background independence. It turns out that this requirement is \emph{vastly} stronger than the habitual Poincar\'e invariance: it suffices to single out a unique representation of $\Al$! More precisely, a fundamental theorem due to Lewandowski, Okolow, Sahlmann, and Thiemann \cite{lost} and Fleishhack \cite{cf} says that \emph{in sharp contrast to Minkowskian field theories,  quantum kinematics of LQG is unique!}

This powerful result in turn leads to a specific quantum Riemannian geometry.  The underlying Hilbert space $\Hkg$ is the space of square integrable functions on the configuration space of connections, with appropriate technical extensions required because of the presence of  the infinite number of degrees of freedom. The construction is mathematically rigorous, with a well-defined, diffeomorphism invariant, regular, Borel measure to define the notion of square integrability --there are no hidden infinities or formal calculations (see, e.g. \cite{al2,almmt1,alrev,ttbook}). Recall that in the familiar Fock space $\F$ in Minkowskian QFTs detailed calculations are facilitated by the decomposition $\F = \oplus_{n} \H_{n}$  of $\F$ into $n$-particle subspaces $\H_{n}$.  There is an analogous decomposition of $\Hkg$ of LQG.  To spell it out, consider graphs $\Gamma$ on $M$, each with a certain number (say $L$) of (oriented) links and a certain number of vertices (say $V$). Consider functions $\Psi_{\Gamma}(A) = \psi (h_{\ell_1},\, \ldots,\,  h_{\ell_L}) $ of connections $A$ which depend on $A$ only though the $L$  Wilson lines  $h_{\ell_{i}}(A) \in \SU(2)$\, (with $i = 1,2\ldots  L$),  \, and are square-integrable with respect to the Haar measure on $[\SU(2)]^{L}$.  They constitute a subspace $\H_{\Gamma}$ of $\Hkg$ (analogous to the $\H_{n}$ for the Fock space) associated with the given graph $\Gamma$.  If one restricts attention to a single graph $\Gamma$, one truncates the theory and focuses only on a finite number of degrees of freedom. This is similar to the truncation in weakly coupled QFTs (such as QED) where the order by order perturbative expansion truncates the theory by allowing only a \emph{finite} number of virtual particles. 

The subspaces $\H_{\Gamma}$ can be further decomposed  into \emph{spin network subspaces} $ \H_{\Gamma,\, j_\ell}\,$ by associating a representation of $\SU(2)$ with each link $\ell$, and tying the incoming and outgoing representations at each vertex with an intertwiner $i_{n}$. However, if a graph $\Gamma_1$ with $L_1$ links is obtained from a larger graph $\Gamma_2$ with $L_2$ links simply by removing $L_2 - L_1$ links, then all the states in $\H_{\Gamma_1}$ can also be realized as states in $H_{\Gamma_2}$ simply by choosing the trivial --i.e., $j_\ell=0$-- representation along each of the additional $L^\prime - L$ links. To remove this redundancy, one introduces the sub-spaces $\H^\prime_\Gamma$ of Hilbert spaces $\H_\Gamma$ by imposing the condition $j_\ell \not= 0$ on every link $\ell$. Then, the total Hilbert space $\Hkg$ can be decomposed as 
\be \label{decomposition} \Hkg = \bigoplus_{\Gamma}\, \H^\prime_{\Gamma} =
\bigoplus_{\Gamma,\, \{j_\ell\} }\, \H_{\Gamma,\, j_\ell}\,   \ee 
where $\{j_\ell\}$ denotes an assignment of a \emph{non-zero} spin label $j_\ell$ with each link $\ell \in \gamma$. Note that each $\H_{\Gamma,\, j_\ell}$ is a \emph{finite} dimensional Hilbert space which can be identified with the space of quantum states of a system of $L$ (non-vanishing) spins. This fact greatly facilitates detailed calculations. The orthonormal basis vectors $|s\rangle := |\Gamma, j_{\ell}, i_{n}\rangle $, defined by a graph $\Gamma$ with an assignment of labels $j_{\ell}, i_{n}$ to its links and nodes,  are called \emph{spin-network states}. They generalize the spin networks originally proposed by Penrose \cite{rp}, where each vertex was restricted to be trivalent. This generalization is essential because states with only trivalent vertices have zero spatial volume \cite{loll1}. We will see in section \ref{s3} that spinfoams provide the quantum transition amplitudes between initial and final spin-network states.

With each regular curve $c$, each regular 2-surface $S$, and each regular 3-dimensional region $R$, there are well defined length, area and volume operators $\h{L}_{c},\, \h{A}_{S}$ and $\h{V}_{R}$ on $\Hkg$, that leave each subspace $\H_{\Gamma}$ invariant, and, thanks to background independence of LQG, their action is quite simple \cite{almmt1}, \cite{loll1}-\cite{eb1}, \cite{alrev,ttbook}. 
For example, the action of $\h{A}_{S}$ on a state $\Psi_{\Gamma}$ is non-trivial only if $S$ intersects at least one link in $\Gamma$ and then the action involves simple $\SU(2)$ operations on the group elements $h_{\ell_{i}}$ associated with links $\ell_{i}$ passing through that point. Similarly, the action of the volume operator $\h{V}_{R}$ is non-trivial only at the nodes of $\Gamma$ and involves simple $\SU(2)$ operations there. Eigenvalues of these operators are discrete. However, the level spacing is not uniform; the levels crowd exponentially as eigenvalues increase, making the continuum limit excellent very rapidly. Of particular interest is the \emph{area gap}, $\underline{\Delta} :=  4\sqrt{3}\pi\gamma\, \lp^2 \,\, $, the smallest non-zero eigenvalue of $\h{A}_{S}$. It serves as the fundamental \emph{microscopic} parameter that then determines the macroscopic parameters --such as the upper bound on matter density and curvature in cosmology-- at which quantum geometry effects dominate.  From the viewpoint of the final quantum theory, area gap is the fundamental \emph{physical} parameter that sets the scale for new LQG effects; it subsumes the mathematical parameter $\gamma$ introduced in the transition from the classical to the quantum theory.
\begin{figure}[]
  \begin{center}
\includegraphics[width=0.5\textwidth]{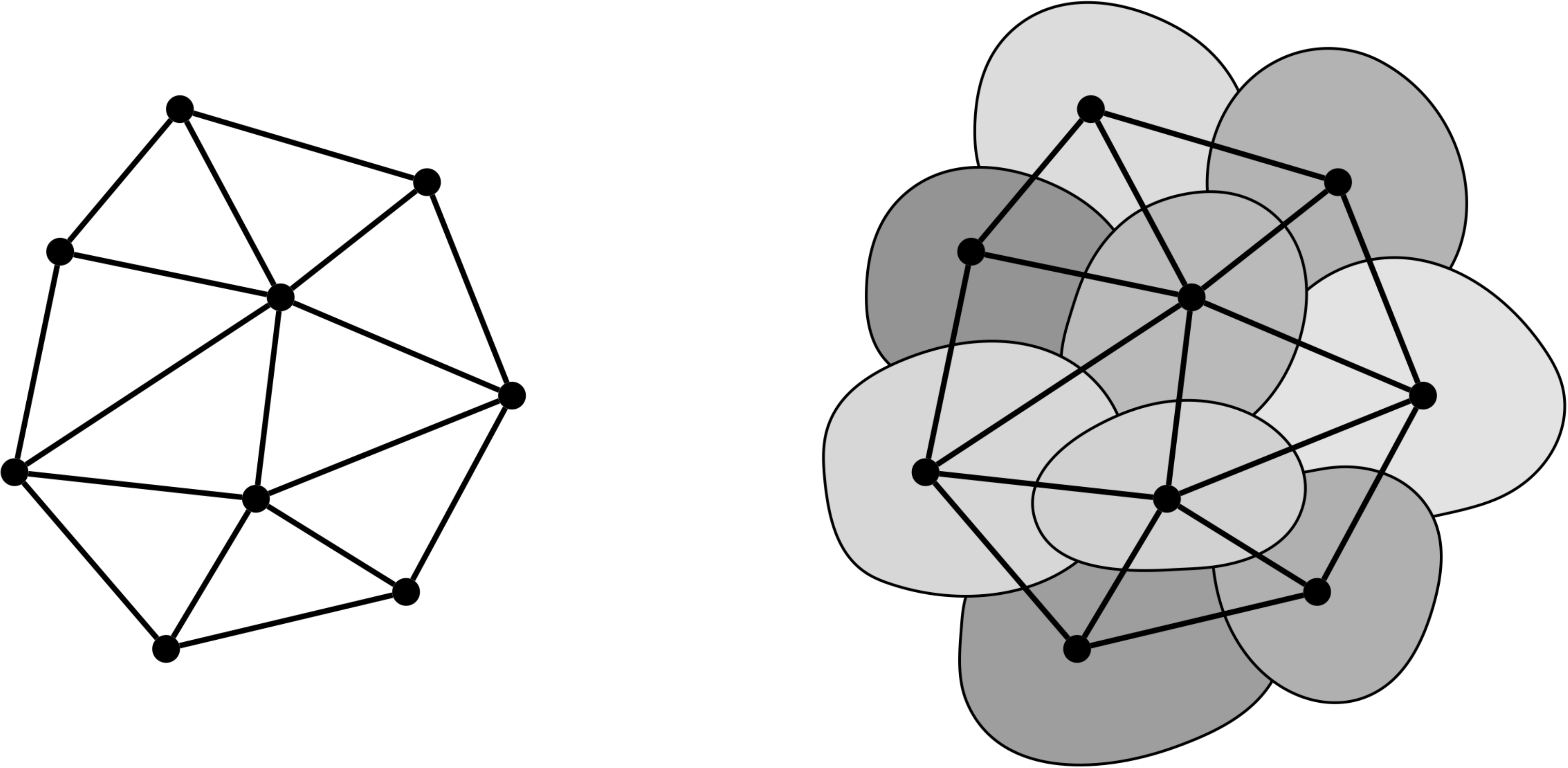}
\caption{Artist's depiction of quanta of geometry. The left figure is a graph $\Gamma$ and the right figure shows $\Gamma$ with its dual cellular decomposition.}
\label{fig:spinnet}
\end{center}
\end{figure}

The simplest way to visualize the elementary quanta of geometry is depicted in Fig.~\ref{fig:spinnet} for a general graph. But for simplicity, let us suppose we are given a $4$-valent graph. Then, one can introduce a dual simplicial decomposition of the $3$-manifold $M$: Each $3$-cell in the decomposition is a topological tetrahedron dual to a node $n$ of $\Gamma$; each face is dual to a link $\ell$. Each $3$-cell can be visualized as an `atom of geometry'. Its volume `resides' at the node, and areas of its faces `reside' at the point at which the face intersects the link of the graph. Thus,  quantum Riemannian geometry is distributional in a precise sense and classical Riemannian structures arise only on coarse graining. Indications of how QFT in curved spacetimes is to emerge from such a quantum geometry can be found in \cite{hstt}.

\section{Quantum dynamics}
\label{s3}

There are two approaches to quantum dynamics, both of which are  rooted in the kinematical framework of section \ref{s2}.  The first is based on Hamiltonian methods and aims at completing the quantization program for constrained systems introduced by Dirac. LQC, discussed in Section \ref{s4}, will illustrate this program and we will also comment on its current status for full LQG in section \ref{s5}. The second approach --that goes under the name of \emph{spinfoams}-- extends the path integral methods used in QFTs but now to a background independent setting. It is well suited to the discussion of field theoretic issues such as the low energy limit of the theory, including $n$-point functions. In this section we summarize the current status of this approach.

\subsection{Spinfoams: General setting and microscopic degrees of freedom}
\label{s3.1}

In elementary quantum mechanics the transition amplitude between an initial and a final state can be computed using the Feynman path integral. Its definition involves three ingredients:  (i) a Hilbert space to specify the initial and the final state, (ii) an action principle, and (iii) a functional integration measure over paths in configuration space. In the case of GR, a path integral formulation was first proposed by Misner \cite{Misner:1957wq} and Wheeler \cite{wheeler} and further developed by Hawking et al \cite{Hawking:1978jz}. The transition amplitude is formally given by a path integral over spacetime geometries $g_{\mu\nu}$,
\begin{equation}
W[q_{ab},q'_{ab}]=\int_{q_{ab}}^{q'_{ab}}\!\!\mathscr{D}[g_{\mu\nu}]\;\mathrm{e}^{\mathrm{i} S[g_{\mu\nu}]/\hbar}\,.
\label{pathint}
\end{equation}	
One assumes that $g_{\mu\nu}$ is a metric of signature $-,+,+,+$ on a $4$-dimensional manifold $\mathcal{M}$ that has initial and final boundaries $M$ and $M'$ with induced $3$-dimensional Riemannian metrics $q_{ab}$ and $q_{ab}'$. As the action $S[g_{\mu\nu}]$ and the functional measure $\mathscr{D}[g_{\mu\nu}]$ are assumed to be invariant under diffeomorphism of $\mathcal{M}$, the transition amplitude $W[q_{ab},q'_{ab}]$ formally provides a solution of the Hamiltonian and $3$-diffeomorphism constraints of quantum geometrodynamics (\ref{admcon}). 
However, while the action of GR can be taken to be the Einstein-Hilbert action, what is missing in (\ref{pathint}) is a definition of the functional measures over spatial geometries $\mathscr{D}[q_{ab}]$ to define the Hilbert space, and  over spacetime geometries $\mathscr{D}[g_{\mu\nu}]$ to define the functional integral. The Hilbert space (\ref{decomposition})  of quantum Riemannian geometries $\Hkg$ in LQG   provides a rigorous definition of the first; \emph{spinfoams} provide a strategy for defining the second \cite{Reisenberger:1996pu}. The starting point is a recasting of the action for GR in terms of gauge fields, as a \emph{topological field theory} with constraints. The topological theory has no local degrees of freedom and is straightforward to quantize. The strategy is to impose the constraints in a controlled way, unfreezing first a finite number of degrees of freedom associated to a cellular decomposition of the manifold, and then defining the full transition amplitude as a limit. We illustrate this strategy below. See \cite{crbook,crfvbook,Perez:2012wv,30years:EB} for detailed reviews. 

\bigskip

A $4$-dimensional topological field theory of the BF type (with the Lorentz group as gauge group) is defined by the action \cite{Baez:1999sr}
\begin{equation}
S_{\mathrm{BF}}[B,\omega]=\int_{\mathcal{M}}B_{IJ}\wedge \mathcal{F}^{IJ}(\omega)\,,
\label{SBF}
\end{equation}
where the indices $I,J\ldots$ refer to the Lie algebra of $SO(1,3)$, $\omega^{IJ}=\omega^{IJ}_\mu(x)dx^\mu$ is a Lorentz connection, $\mathcal{F}^{IJ}=d\omega^{IJ}+\omega^I{}_{K}\wedge \omega^{KJ}$ its curvature and $B^{IJ}=B^{IJ}_{\mu\nu}(x)\,dx^\mu\wedge dx^\nu$ a two-form with values in the adjoint representation. Following the conventions generally used in the spinfoam literature, we work with forms and  generally do not display the spacetime manifold indices  $\mu, \nu \ldots$ (in contrast to the usual conventions in the Hamiltonian theory). The $SO(1,3)$ Cartan-Killing metric $\eta_{IJ}$  enables one to freely raise and lower the internal indices, and the alternating tensor is denoted $\epsilon_{IJKL}$.  As in GR, the action (\ref{SBF}) is invariant under diffeomorphisms of $\mathcal{M}$ and Lorentz gauge transformations. However, compared to GR, the topological theory has a much larger symmetry group: The action (\ref{SBF}) is also invariant under shifts of the $B$ field of the form
\begin{equation}
B^{IJ}\;\rightarrow \; B^{IJ}\;+\;\mathcal{D} \Lambda^{IJ}\,.
\label{Bshift}
\end{equation}
where $\mathcal{D}\Lambda^{IJ}=d\Lambda^{IJ}+\omega^I{}_K\wedge\Lambda^{KJ}+\omega^J{}_K\wedge\Lambda^{KI}$ is the covariant derivative of a one-form $\Lambda^{IJ}=\Lambda^{IJ}_\mu(x)dx^\mu$. It is this symmetry that results in \emph{topological invariance} and the absence of local degrees of freedom. At the classical level, i.e., requiring the stationarity of the action with respect to variations $\delta B$ and $\delta \omega$, we obtain the equations of motion
\begin{equation}
\mathcal{F}^{IJ}(\omega)=0\,\quad\textrm{and}\quad \mathcal{D}B^{IJ}=0\,.
\label{EqBF}
\end{equation} 
The first equation tells us that the Lorentz connection $\omega$ is  flat, and therefore locally can be written as a pure gauge configuration. The second equation, together with the invariance (\ref{Bshift}), tells us that the $B$ field can be written locally as the covariant derivative of a one-form $\Lambda^{IJ}$, i.e., $B^{IJ}=\mathcal{D}\Lambda^{IJ}$, and locally all solutions of the equations of motion are equal modulo gauge transformations. The only dynamical degrees of freedom of the theory have a global nature and capture the topological invariants of the manifold. While this description is in terms of classical equations of motion, the conclusion that there are no local degrees of freedom holds also at the quantum level \cite{Horowitz:1989ng}-\cite{Atiyah:1989vu}.

General relativity can be formulated in the same language as the topological theory described above, introducing the Lorentz group $SO(1,3)$ as internal gauge group and adopting Einstein-Cartan variables as fundamental variables: a Lorentz connection $\omega^{IJ}=\omega^{IJ}_\mu(x)dx^\mu$ and a coframe field $e^I=e_\mu^I(x)dx^\mu$. The spacetime metric is a derived quantity given by $g_{\mu\nu}(x)=\eta_{IJ}\,e_\mu^I(x) e_\nu^J(x)$. In these variables, the action for GR takes the form 
\begin{equation}
S_{\mathrm{GR}}[e,\omega]=\frac{1}{16\pi G}\int_{\mathcal{M}}\;\Big(\,\frac{1}{2}\epsilon_{IJKL}e^K\wedge e^L -\;\frac{1}{\gamma}e_I\wedge e_J\Big)\wedge \mathcal{F}^{IJ}(\omega)\;,
\label{SGR}
\end{equation}
where we have included a topological term with a coupling constant $\gamma$ coinciding with the Barbero-Immirzi parameter encountered in the canonical theory (Sec.~\ref{s2.2}) \cite{Holst:1995pc,Wieland:2010ec}. As with the action (\ref{SBF}), the theory is invariant under diffeomorphisms of $\mathcal{M}$ and under Lorentz gauge transformations. The difference is that now there is no analogue of the topological symmetry (\ref{Bshift}): The theory has infinitely many dynamical degrees of freedom, two per point, and the equations of motion are non trivial,
\begin{equation}
e^I\wedge \mathcal{D}e^J=0
\,\quad\textrm{and}\quad
\epsilon_{IJKL}\,e^J\wedge \mathcal{F}^{KL}(\omega)=0  \,.
\label{EinsteinEqs}
\end{equation}
The first equation is the vanishing condition for the torsion $T^I(e,\omega)=\mathcal{D}e^I$ and, when this condition is satisfied, the second equation is equivalent to the vacuum Einstein equations. Note that the Barbero-Immirzi parameter $\gamma$ does not appear in the classical equations of motions. 
\bigskip

The key observation in the formulation of spinfoams is that  GR with action (\ref{SGR}) can be understood as a topological theory with action (\ref{SBF}),  together with the requirement that there exists a one-form $e^I$ such that the $B$ field takes the form \cite{Baez:1994zz}:
\begin{equation}
B_{IJ}\, =\,
\frac{1}{16\pi G}\Big(\,\frac{1}{2}\epsilon_{IJKL}\,e^K\wedge e^L\,-\,\frac{1}{\gamma}e_I\wedge e_J\Big)\,.
\label{Bconstraint}
\end{equation}
This condition can be imposed as a constraint in the action \cite{Plebanski:1977zz}-\cite{Reisenberger:1996ib}. 
Recall that a two-form $\Sigma$ is said to be \emph{simple} if it can be written as the exterior product of two one-forms, i.e., $\Sigma=\eta\wedge\theta$. The constraint (\ref{Bconstraint}) requires that the $B$ field is `$\gamma$-simple' and is called \emph{simplicity constraint}. In (\ref{SBF}), the $B$ field plays the role of Lagrange multiplier for the curvature $\mathcal{F}$ and imposes that it must vanish;  the constraint (\ref{Bconstraint}) on $B$ frees $\mathcal{F}$ and allows non-flat connections. Moreover, this constraint  breaks the topological symmetry (\ref{Bshift}) and allows one to introduce a metric $g_{\mu\nu}(x)=\eta_{IJ}\,e^I_\mu(x)e^J_\nu(x)$. Imposing the constraint (\ref{Bconstraint}) everywhere on the $4$-manifold, unfreezes infinitely many degrees of freedom (two per point) and recovers full GR. On the other hand, if the constraint is imposed only on a finite ``skeleton'' of the $4$-manifold $\mathcal{M}$, then we unfreeze only a finite number of degrees of freedom and a truncation of GR is obtained. A classical spinfoam model is a topological field theory of the type (\ref{SBF}) with a finite number of dynamical degrees freedom associated to a network of topological defects \cite{Bianchi:2009tj}. The defects are introduced by equipping the $4$-manifold $\mathcal{M}$ with a cellular decomposition.

A cellular decomposition is a way to present a manifold as composed of simple elementary pieces, cells with the topology of a ball. The simplest example is a triangulation, the decomposition of a manifold into $4$-simplices, tetrahedra, triangles, segments and points as used in Regge calculus \cite{Regge:1961px}. In spinfoams we consider decompositions that allow more general cells \cite{Oeckl:2005rh}. We denote $\Delta_n$ the set of cells of dimension $n$ and say that two $n$-cells are adjacent if they share a $(n-1)$-cell on their boundary. A $4$-manifold $\mathcal{M}$ equipped with a cellular decomposition $\mathcal{M}_\Delta=\Delta_4\cup\Delta_3\cup\Delta_2\cup\Delta_1\cup\Delta_0$ is called a cellular manifold. It is also useful to introduce the notion of $2$-skeleton $\mathcal{S}_2=\Delta_2\cup\Delta_1\cup\Delta_0$ of the cellular manifold $\mathcal{M}_\Delta$. The role of the $2$-skeleton $\mathcal{S}_2$ is two-fold: First, as $\mathcal{S}_2$ is a branched surface embedded in $\mathcal{M}$, it is immediate to impose the simplicity constraint (\ref{Bconstraint}) on each of its elements; this constraint unfreezes the curvature $\mathcal{F}(\omega)$ on the $2$-skeleton only, therefore turning $\mathcal{S}_2$ into a network of topological defects where the curvature is supported. 
Second, the $4$-manifold $\mathcal{M}'=\mathcal{M}-\mathcal{S}_2$ is path-connected but not simply-connected: there are non-contractible closed paths in $\mathcal{M}'$ that encircle elements of the $2$-skeleton. As a result, the Wilson lines --or holonomies-- of the connection $\omega$ that encircle the topological defects are non-trivial. This is how the \emph{loops} of LQG \cite{loops1,loops2}, and the 4-dimensional version of the holonomy $h_{\ell}(A)$ (\ref{hAE}), arise in the spinfoam dynamics.

\begin{figure}[]
  \begin{center}
\includegraphics[width=0.4\textwidth]{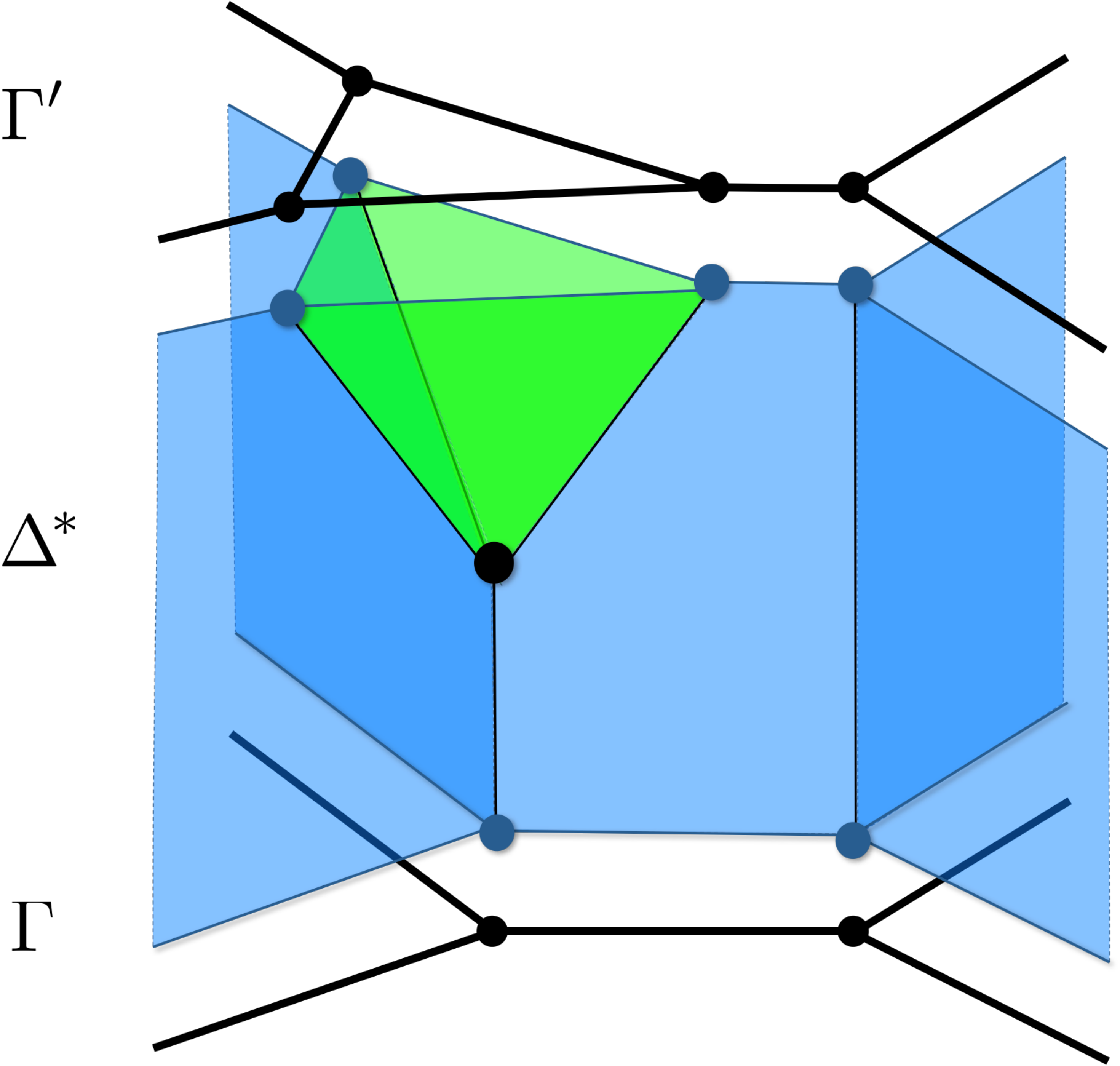}
\caption{Depiction of a $2$-complex (or \emph{spinfoam}) that interpolates between an initial and a final graph (or \emph{spin network}).}
\label{fig:spinfoam}
\end{center}
\end{figure}

The microscopic degrees of freedom of the theory are best described in terms of an abstract $2$-complex $\mathcal{C}_2$ which captures the homotopy group $\pi_1(\mathcal{M}-\mathcal{S}_2)$ of the manifold with topological defects and non-trivial holonomies. $\mathcal{C}_2$ is defined  by introducing a dual cellular decomposition of the $4$-manifold $\mathcal{M}_{\Delta^*}=\mathcal{B}_4\cup\mathcal{B}_3\cup f\cup e\cup v$ with $n$-dimensional cells such that there is a \emph{vertex} $v$ in $\mathcal{C}_2$ per $4$-cell $\Delta_4$, an \emph{edge} $e$ per $3$-cell $\Delta_3$ and a \emph{face} $f$ per $2$-cell $\Delta_2$ of $\mathcal{M}_\Delta$. Two vertices are connected by an edge if they are dual to two adjacent $4$-cells. This abstract $2$-complex $\mathcal{C}_2=f\cup e\cup v$ --also called a \emph{spinfoam}-- is the set of faces, edges and vertices in $\Delta^*$, together with their adjacency conditions \cite{Baez:1999sr}. Non-contractible loops in $\mathcal{M}'=\mathcal{M}-\mathcal{S}_2$ correspond to cyclic sequences of edges that bound a face $f$ of  $\mathcal{C}_2$ that is dual to a $2$-cell in $\mathcal{S}_2$. Moreover a foliation of the manifold $\mathcal{M}=M\times\mathbb{R}$ corresponds to a slicing for the $2$-complex $\mathcal{C}_2$ into graphs $\Gamma$ with a link $\ell$ for each intersected face and a node $n$ for each intersected edge of the $2$-complex (See Fig.~\ref{fig:spinnet} and \ref{fig:spinfoam}). These are the graphs that were introduced in the discussion of spin networks in section \ref{s2.2}.

Up to this point, the construction is classical and defines a truncation of GR with a finite number of degrees of freedom. It is a field-theoretic truncation in the sense that we have \emph{not} discretized derivatives as it is done for instance in lattice field theory: our variables are still fields $B_{IJ}$ and $\omega^{IJ}$ but --apart from a finite number of dynamical degrees of freedom that capture the non-trivial topology of $\mathcal{M}-\mathcal{S}_2$ -- they are pure gauge. As a result, one can determine their functional integration measure $\mathscr{D}[B_{IJ}]\mathscr{D}[\omega^{IJ}]$ rigorously as it is done for topological QFTs \cite{Horowitz:1989ng}-\cite{Atiyah:1989vu}.

The key step that leads one from a topological field theory to one with true degrees of freedom is of course the imposition of the $\gamma$-simplicity contraint (\ref{Bconstraint}).  The Engle-Pereira-Rovelli-Livine (\emph{EPRL}) spinfoam model \cite{Engle:2007qf}-\cite{Engle:2007wy},
and its extension to general cellular decompositions \cite{Kaminski:2009fm}-\cite{Thiemann:2013lka},
provides a specific implementation of this step. Inserting a resolution of the identity in the path integral using the spin-network basis (\ref{decomposition}) of the Hilbert space $\H_\Gamma$ of LQG, one can write the transition amplitude from an initial spin-network state $|s\rangle=|\Gamma,j_\ell,i_n\rangle$ and the final spin-network state $|s'\rangle=|\Gamma',j'_\ell,i'_n\rangle$ in the combinatorial form
\begin{equation}
W_\Delta[s,s']=\sum_{j_f, i_e}\prod_{f\in \Delta^*}A_f(j_f)\prod_{v\in \Delta^*} A^{(\gamma)}_v(j_f,i_e)\,,
\label{Wspinfoam}
\end{equation}
where $\Delta$ is a cellular decomposition with dual complex $\Delta^*$ that interpolates between the graphs $\Gamma$ and $\Gamma'$ (see Fig.~\ref{fig:spinfoam}).  The face amplitude $A_f(j_f)$ and the vertex amplitude $A^{(\gamma)}_v(j_f,i_e)$ fully encode the dynamics of the theory truncated to the decomposition $\Delta$ and provide a definition of the path integral over truncated spacetime geometries (\ref{pathint}). The EPRL model provides both these amplitudes. The vertex amplitude $A^{(\gamma)}_v(j_f,i_e)$ is an invariant built out of $\gamma$-simple representation of the Lorentz group $SO(3,1)$ \cite{Martin-Dussaud:2019ypf}. Its form is analogous to the one of the $\{6j\}$ symbol encountered in the `composition of angular momenta' in quantum mechanics and in $3d$ quantum gravity \cite{Ponzano}. 

For a fixed cellular decomposition $\Delta$, the spinfoam path integral has only a finite (although large) number of degrees of freedom. There are no ultraviolet divergencies because of the discrete nature of the sum over $SU(2)$ representations which reflects the discreteness of quantum Riemannian geometry and the area gap described in Sec.~\ref{s2.2}. In the presence of a positive cosmological constant, the spinfoam amplitude $W_\Delta[s,s']$ at fixed $\Delta$ is also infrared finite \cite{Han:2010pz}-\cite{Haggard:2014xoa} 
 as the  $q$-deformation of the gauge group results is a maximum spin $j_\mathrm{max}$ and a physical cut-off for large-volume bubbles in the dual complex \cite{Riello:2013bzw,Dona:2018pxq}.

The full spinfoam dynamics with infinitely many degrees of freedom is formally given by a sum over decompositions $W[s,s']=\sum_{\Delta:\Gamma\to\Gamma'}W_\Delta[s,s']$. A key open issue is the mathematical definition of this sum over $\Delta$.  Group field theory \cite{DePietri:1999bx,Oriti:2013aqa} provides a Feynman diagrammatic scheme for summing over decompositions as a perturbative expansion in the number of spinfoam vertices. We note that the definition of the infinite sum over decomposition has to satisfy a number of non-trivial consistency conditions. First, redundancies arise because of cellular decompositions related by refinement. These redundancies are analogous to the ones discussed in the definition of the Hilbert space $\Hkg$ (\ref{decomposition}) and can be treated similarly \cite{Kaminski:2009fm,Kaminski:2011bf}, \cite{Rovelli:2010qx}. Second, while $W[s,s']$ is not required to be finite, it must satisfy the consistency conditions for the definition of a physical Hilbert space $\Hpg$ \cite{Thiemann:2013lka}. With these caveats, the physical scalar product of two states is given by $\langle\psi|\hat{W}|\psi'\rangle=\sum_{s,s'} \psi(s)W[s,s'] \psi'(s')$, where $\psi(s)$ and $\psi'(s)$ are superpositions of spin network states that define physical states.

\subsection{Spinfoams: Reconstructing a semiclassical spacetime}
\label{s3.2}

Besides providing a covariant definition of the dynamics, the path integral over spacetime geometries and its proposed spinfoam realization (\ref{Wspinfoam}) provide also a bridge to the reconstruction of a classical spacetime with small quantum fluctuations over it. Formally, in the limit $\hbar\to 0$, one can approximate the path-integral (\ref{pathint}) by perturbations around a saddle-point: A classical spacetime with metric $\bar{g}_{\mu\nu}$ is given by a saddle point of the action $S[g_{\mu\nu}]$, and quantum perturbations $\delta g_{\mu\nu}$ over this background --\emph{gravitons}-- are described by an effective field theory with action $S[\bar{g}_{\mu\nu}+\delta g_{\mu\nu}]$ \cite{Donoghue:1994dn,Burgess:2003jk}. On the other hand, LQG is a \emph{background-independent} theory as it does not involve a choice of the classical background in its formulation. The idea is to  identify a semiclassical regime of LQG, where GR and QFT on a curved spacetime are recovered,  by introducing semiclassical states that have the background $\bar{g}_{\mu\nu}=\langle g_{\mu\nu}\rangle$ as expectation value. Then one can investigate $n$-point correlation functions of observables of the quantum geometry in this state \cite{Rovelli:2005yj,Bianchi:2006uf}. We describe this strategy below.

The first step is to introduce a semiclassical state of the quantum geometry \cite{Thiemann:2000bw}-\cite{Bianchi:2016hmk}. 
Let us consider a graph $\Gamma$. A spin-network basis state $|\Gamma,j_\ell,i_n\rangle$ describes a quantum Riemannian geometry: It is a simultaneous eigenstate of the volume of $3$-cells dual to each node of $\Gamma$ and of the area of each face shared by two adjacent $3$-cells. While volumes and areas have definite value, because of the non-commutativity of quantum geometry (and in particular the non-commutativity of dihedral angles between faces), Heisenberg uncertainty relations arise and the shape of each $3$-cell is fuzzy. A semiclassical state for each \emph{individual} $3$-cell can be obtained by considering a \emph{coherent} superposition of volume eigenstates $i_n$ that is peaked on the shape of a Euclidean polyhedron \cite{Bianchi:2010gc, Bianchi:2011ub}. One then obtains a `many-body' state that describes an un-entangled collection of semiclassical polyhedra. Note that the area of faces of adjacent polyhedra match by construction, but the shape of the faces does not: the geometry is \emph{twisted} \cite{Dittrich:2008ar,Freidel:2010aq}. Moreover, as the polyhedra have faces with a definite area, they have maximal dispersion in the conjugate variable --the extrinsic curvature. A semiclassical state on the graph $\Gamma$ is an entangled state of semiclassical polyhedra that is peaked on a truncation of the intrinsic and the extrinsic geometry of space --the geometrodynamical pair $(q_{ab},p^{ab})$ of Sec.~\ref{s2.1}-- and has long-range correlations \cite{Bianchi:2012ev,Baytas:2018wjd}. The description up to this point is kinematical. It is the dynamics (in its canonical or spinfoam formulation) that determines the allowed pairs $(q_{ab},p^{ab})$ and the specific correlations.

Once semiclassical states $|\psi\rangle$ and $|\psi'\rangle$ for the initial and the final state have been selected (each of the form $\sum_{j_\ell, i_n} c_{j_\ell,i_n}\,|\Gamma,j_\ell,i_n\rangle $), one could
define correlation functions for geometric observables  $\mathcal{O}_i$ as
\begin{equation}
G_{ij}=\frac{\langle \psi'|\mathcal{O}_i\,\hat{W}_\Delta\,\mathcal{O}_j|\psi\rangle}{\langle \psi'|\hat{W}_\Delta|\psi\rangle}\,,
\label{Gij}
\end{equation}
where $\hat{W}_\Delta$ is the spinfoam (\ref{Wspinfoam}) seen as an operator. Note that there is a non-trivial consistency condition that ties the initial and the final state: the initial semiclassical state $|\psi\rangle$ is peaked on a canonical pair $(\bar{q}_{ab},\bar{p}^{ab})$ that, via the dynamics, determines a background spacetime geometry $\bar{g}_{\mu\nu}$; the final semiclassical state $|\psi'\rangle$ is required to determine the same classical background geometry so that (\ref{Gij}) represents correlations of perturbations $\delta g_{\mu\nu}$ over the background $\bar{g}_{\mu\nu}$. Moreover, this formulation requires that we prescribe the semiclassical state for all of spacetime (together with its asymptotic structure), while the correlation function $G_{ij}$ probes only a finite spacetime region. These two technical difficulties are generally addressed by adopting the boundary amplitude formalism \cite{Oeckl:2005bv,Bianchi:2006uf}: one considers a finite spacetime region together with its cellular decomposition $\Delta$ with boundary. Instead of having an initial and a final state, one then has a single boundary state $|\Psi\rangle$ and the spinfoam provides a linear functional $\langle W_\Delta|$ over the space of boundary states. The correlation function is now given by the formula $G_{ij}=\langle W_\Delta| \mathcal{O}_i \mathcal{O}_j|\Psi\rangle/\langle W_\Delta|\Psi\rangle$. Besides providing a computational technique, the boundary amplitude formalism also allows us to address a key conceptual difficulty in the definition of $n$-point correlation functions in a diffeomorphism invariant theory mentioned in section \ref{s1}. Formally, one might define a quantum-gravity correlation function $G(x,y)$ by inserting local operators $\mathcal{O}(x) \mathcal{O}(y)$ in the path integral (\ref{pathint}). However, as the action and the measure is invariant under diffeomorphisms that send the point $x$ into $\tilde{x}$, we would have that the correlation function is also invariant, $G(\tilde{x},\tilde{y})=G(x,y)$, and therefore constant. How do we recover the typical $1/d^2$ behavior of correlation functions of QFT? The choice of boundary semiclassical state $|\Psi\rangle$ provides an average geometry $\bar{g}_{\mu\nu}$ with respect to which to measure the geodesic distance $d$ and, as the points $x$ and $y$ now belong to the same boundary, this allows us to anchor the points to the boundary geometry and determine $d$ before computing the correlation function \cite{Bianchi:2006uf}.

The strategy described above is used in detailed computations of correlation functions in spinfoams, but so far by restricting attention to a decomposition of the simplest kind. For a region including a single $4$-cell with the topology of a $4$-simplex, the EPRL vertex amplitude  $A^{(\gamma)}_v(j_f,i_e)$ together  with a boundary state $|\Psi\rangle$ peaked on a triangulation is shown to reproduce the exponential of the action of Regge's discretization of GR, $\langle W_\Delta|\Psi\rangle\sim \mathrm{e}^{\,\mathrm{i}S_{GR}/\hbar}+\mathrm{cc}$. This result, derived in a saddle-point approximation \cite{Barrett:2009gg,Barrett:2009mw} and tested numerically \cite{Dona:2019dkf}, provides a $4$-dimensional Lorentzian generalization of the classic Ponzano-Regge formula for $3d$ quantum gravity \cite{Ponzano}. Moreover, correlation functions for geometric operators such as areas and dihedral angles have been computed and shown to coincide with the correlation functions of perturbative quantum gravity in the Regge truncation \cite{Bianchi:2009ri,Bianchi:2011hp}. There has also been growing interest in direct tests of how curvature arises on the interior of the $2$-skeleton of simple cellular decompositions \cite{Bonzom:2009hw}-\cite{Dona:2020tvv}. These results provide a first step in the calculation of correlations at fixed cellular decomposition and in the exploration of the effects of a sum over decompositions. Recently developed methods for effective analytical \cite{Asante:2020qpa,Asante:2020iwm} and numerical computations \cite{Bahr:2015gxa}-\cite{Han:2020npv} 
for larger cellular decompositions are providing new tools for addressing the conceptual issues of the reconstruction of a smooth semiclassical spacetime with long-range correlations.

\section{Loop Quantum Cosmology}
\label{s4}

In this section we switch gears to applications and illustrate how salient features of the quantum Riemannian geometry lead to unforeseen and exciting possibilities in the investigation of the early universe. While there are also other contexts in which LQG has led to new insights, we chose this example because it has drawn the most attention within the community so far.

Friedman, Lema\^{i}tre, Robertson, Walker (FLRW) solutions of GR, have a big bang singularity if matter satisfies the standard energy conditions. However, already in the 1970s Wheeler expressed the hope that quantum gravity effects would resolve this  singularity and there has been considerable work in quantum cosmology since then. In LQC, Wheeler's hope has been realized in a precise sense: the big bang is replaced by a specific big bounce and all physical observables remain finite throughout their evolution. Therefore one can extend the standard inflationary scenario to the deep Planck regime in a self-consistent manner, leading to observable predictions.  It turns out that, thanks to an unforeseen interplay between the ultraviolet and the infrared, the quantum geometry effects from the pre-inflationary phase of dynamics leave certain signatures at the \emph{largest} angular scales that can account for certain anomalies observed in the cosmic microwave background (CMB). We will first summarize results on singularity resolution and then turn to the interplay between fundamental theory and observations. 

\subsection{The big bounce of LQC}
\label{s4.1}

Big Bang is often heralded as the clear-cut beginning of our physical universe. However,  as Einstein himself pointed out, it is a prediction of GR in a regime that is outside its domain of validity: {\sl ``One may not assume the validity of field equations at very high density of field and matter and one may not conclude that the beginning of the expansion should be a  singularity in the mathematical sense''} \cite{einstein1946}.  Indeed, we know that quantum effects dominate in neutron stars because of high density  $\rho \sim 10^{18}\, \mathrm{kg}/\mathrm{m}^3$; without the Fermi degeneracy pressure, neutron stars would not even exist!  Similarly, gravity effects are expected to  dominate in the Planck regime  --i.e., once matter density reaches $\rho \sim 10^{97}\, \mathrm{kg}/\mathrm{m}^3$--  and qualitative change the classical GR dynamics, well before the big bang is reached.  In fact, when cosmologists now speak of the `big bang' they generally refer to a hot phase in the early universe (e.g., at the end the reheating process after inflation); not the initial singularity in the FLRW models! (See, e.g., \cite{halper}.)  By now, resolution of the big bang singularity has been arrived at in a variety of programs. However, it is fair to say that the systematic conceptual and mathematical framework was first introduced in detail in LQC (see, e.g., \cite{mb}-\cite{kp}, \cite{asrev,30years:IAPS}).

We will now summarize the main ideas and illustrate the key results. Standard investigations of the early universe are carried out assuming that spacetime is well approximated by a spatially flat FLRW background spacetime, together with first order cosmological perturbations, described by quantum fields. Therefore, as in every approach to quantum cosmology, in LQC one starts with this cosmological sector of GR. 
The classical FLRW spacetime --that is characterized by  a scale factor $a(t)$ together with matter fields, say $\phi(t)$-- is now replaced by a quantum state $\Psi(a,\phi)$ that is to satisfy the \emph{quantum versions} of the Friedmann and Raychaudhuri equations.  Note that reference to the proper time $t$ has disappeared --quantum dynamics is relational, \`a la Leibnitz:  for example, one can use the matter field $\phi$ as an internal clock, and describe how the scale factor evolves with respect to it. Quantum fields representing cosmological perturbations now propagate on a \emph{quantum} geometry $\Psi(a,\phi)$.

There are two features of LQC that distinguish it from the older Wheeler-DeWitt theory, i.e., cosmological models in the framework of quantum geometrodynamics:  (i)  mathematical precision and conceptual completeness of the underlying framework, which in turn, led to  (ii) a singularity resolution through a quantum bounce with specific physical attributes. The starting point is the LQG quantum kinematics,  summarized in section \ref{s2.2}, but now suitably restricted by the requirements of spatial homogeneity and isotropy. Thus, there is a symmetry-reduced holonomy-flux algebra $\Alred$ where the links $\ell$, surfaces $S$ and test fields $f^{i}$ are now restricted by the underlying symmetry. It turns out that there is a `residual' group of diffeomorphisms on the spatial 3-manifold $M$ that has non-trivial action on $\Alred$. Therefore, as in full LQG, one can again use the requirement of background independence to demand invariance under this action and select a unique representation of $\Alred$ \cite{ac,eht}. As with $\Hkg$ in full LQG, 
the Hilbert space $\Hkr$ carrying the representation of $\Alred$ has novel features that descend from the area gap $\underline{\Delta}$ of LQG (which are not shared by the Schr\"odinger representation normally used in quantum geometrodynamics). As a result, the quantum version of the Hamiltonian constraint is also strikingly different from the Wheeler-DeWitt equation of quantum geometrodynamics.  One can now take a quantum state $\Psi(a, \phi)$ that is sharply peaked on the classical trajectory in the `late epoch'  --when spacetime curvature and matter density are low compared to the Planck scale-- and use the quantum Hamiltonian constraint  to evolve it back in time (w.r.t. the `matter clock') towards higher curvature and density.  Interestingly, the wave packet follows the classical trajectory till the density  increases to $\rho\, \sim 10^{-4} \rhop$. Then the quantum geometry effects cease to be negligible  and the evolution departs from the classical trajectory. $\Psi(a,\phi)$  is still sharply peaked but the quantum corrected trajectory its peak now follows  undergoes a bounce when the density reaches  a critical, maximum value $\rcr  := {18\pi G\hbar^{2}}/( \underline{\Delta}^{3}) \, \approx 0.41 \rhop$, and then it starts decreasing. Again when the density falls to $\rho \sim 10^{-4} \rhop$, quantum corrections become negligible and GR is again an excellent approximation (see, e.g., \cite{aps,asrev,30years:IAPS}). Thus, quantum geometry effects create a bridge joining our expanding FLRW branch to a contracting FLRW branch in the past. These qualitatively new features arise without having to introduce matter that violates any of the standard energy conditions, and without having to introduce new boundary conditions, such as the Hartle-Hawking `no-boundary proposal'; they are consequences just of the quantum corrected Einstein's equations. In particular,  in the limit $\underline{\Delta} \to 0$ --i.e., in which we ignore quantum geometry effects--  $\rcr \to \infty$.  Thus, the existence of the bounce and the upper bound on the density (and curvature) can be directly traced back to quintessential features of quantum geometry. In the models that have been worked out in detail also in the Wheeler DeWitt theory,  there is no bounce, and both matter density and curvature remain unbounded above \cite{aps}.

Many of the consequences of the LQC dynamics can be readily understood in terms of \emph{effective equations} that capture the qualitative behavior of sharply peaked $\Psi(a,\phi)$. They encapsulate the leading order corrections to the classical Einstein's equation in the Planck regime. While these corrections modify the geometrical part (i.e. left side) of Einstein's equations, it is convenient to move them to the right side by a mathematical rearrangement. Then, the quantum corrected Friedmann equation assumes the form
\be \Big(\frac{\dot{a}}{a} \Big)^2 = \frac{8\pi G\,\rho} {3}\, \left(1 -  \frac{{\rho}}{\rcr} \right) \, . \ee
where the second term on the right side represents the quantum correction. Without this term, i.e., in classical GR,  the right side is positive definite, whence $\dot{a}$ cannot vanish: the universe either expands out from the big bang or contracts into a big crunch. But, with the quantum correction, the right side vanishes at $\rho =\rcr$,  whence $\dot{a}$ vanishes there and
the universe bounces. This occurs only  because the LQC correction $\rho/\rcr$ \emph{naturally} comes with a \emph{negative} sign which gives rise to an effective `repulsive force' in the Planck regime. The occurrence of this negative sign is non-trivial: in the standard brane-world scenario,
for example, Friedmann equation is also receives a $\rho/\rcr$ correction but it comes with a positive sign (unless one makes the brane tension negative by hand) whence the singularity is not resolved naturally. Finally, there is an excellent match between analytical results within the quantum theory, numerical simulations and effective equations. Finally, these considerations have been extended to include spatial curvature, non-zero cosmological constant, anisotropies (see, e.g., 
\cite{asrev,30years:IAPS} and references therein) and also the simplest inhomogeneities captured by the Gowdy models \cite{gowdy} and to the Brans-Dicke theory \cite{bdtheory}.

\noindent{\it Remarks:}\\
(i) Note that, as we explained above, the term ``effective equations" 
refers to the leading order LQC corrections to Einstein's equations for states $\Psi (a,\phi)$ that are sharply peaked on a classical trajectory at late times. In contrast to the procedure used in standard effective field theories, one does not integrate out ultraviolet modes.\\
(ii) The big bounce has been analyzed in detail in a large number of LQC papers, using Hamiltonian, cosmological-spinfoam and `consistent histories' frameworks (see, e.g., \cite{asrev,30years:IAPS}). Taken together, these results bring out the robustness of the main results. Recently some concerns were expressed about certain technical points \cite{mb2}. Their main thrust was already addressed, e.g., in \cite{cs1,kp,asrev} and in the Appendix of \cite{aa-grg}. However, one should keep in mind that, as in all approaches to quantum cosmology, in LQC the starting point is the symmetry reduced, cosmological sector of GR. The issue of arriving at LQC from LQG remains a subject of active investigation (see section \ref{s5}).

\subsection{Can one see quantum geometry effects in the sky?}
\label{s4.2}

The quantum corrected dynamics of LQC has been used to make contact with observations 
especially (but not exclusively) in the context of inflation. This paradigm assumes that, in its early history, the universe underwent a nearly exponential expansion via classical GR equations, in response to a slow roll of a scalar field down a suitable potential. This is taken to be the background space-time geometry on which quantum fields representing cosmological perturbations evolve. Now, an exponential  expansion corresponds to a deSitter metric and, thanks to its maximum symmetry, linear quantum fields on de Sitter admit a preferred state called the Bunch-Davies (BD) vacuum. Therefore, one assumes that the cosmological perturbations were in the BD vacuum at the onset of the slow roll and calculates the temperature-temperature (TT) power spectrum at the end of inflation, which turns out to be  nearly scale invariant. This motivates a standard ansatz ({\it SA})  for the  \emph{primordial power spectrum} featuring 2-parameters, the amplitude $A_{s}$ and the spectral index $n_{s}$:\, $\mathcal{P}(k) = A_{s}\, \left(\f{k}{k_{\star}}\right)^{n_{s}-1}\!\!\!$  where $k$ is as usual the wave number and $k_{\star}$ a fiducial value.  (Thus, $n_{s} =1$ would correspond to exact scale invariance.) 

To make contact with observations, cosmologists use the following procedure.  The primordial power spectrum is then time-evolved using known (astro)physics and leads to 3 other power spectra (featuring `electric polarization' and the `lensing potential')  that can be observed in the CMB. This procedure requires an input of 4 additional parameters, usually taken to be the baryonic and cold matter densities $\Omega_{b}h^{2}$ and $\Omega_{c}h^{2}$, (that are key to the propagation of perturbations starting form the end of inflation), and the optical depth $\tau$ and the angular scale associated with acoustic oscillations $100\theta_{\star}$ (that are key to the  propagation to the future of the CMB surface). By fitting the 4 predicted power spectra with observations, one determines the best fit values (and standard deviations) for the 6 cosmological parameters \cite{planck6}. This is then the 6-parameter $\Lambda$CDM universe that best describes the large scale structure of our universe! One can now compute other observables for this model universe and compare those predictions with observations as additional checks on the model. By and large, the standard model provided by the PLANCK collaboration agrees extremely well with observations \cite{planck6}. However, there are also certain `anomalies'. Presence of any one of these anomalies is not statistically significant. However, taken together, two of them already imply that according to standard inflation we live in an exceptional universe that occurs only in one in $\sim 10^{6}$ realizations of the posterior probability distribution.

One can regard these anomalies as opportunities to discover new physics. As the PLANCK collaboration  put it, \cite{planck1}  {\sl ``...if any anomalies have primordial origin, then their large scale nature would suggest an explanation rooted in fundamental physics. Thus it is worth exploring any models that might explain an anomaly (even better, multiple anomalies) naturally, or with very few parameters.''}  Note that the burden on the potential new physics is enormous because it has to resolve the tension caused by the anomalies, \emph{without affecting the very large number of predictions that agree with observations.} Yet, this is just what happens in LQC: several of these anomalies can be alleviated by the quantum geometry effects in the pre-inflationary history of the universe, leaving the successes of standard inflation in tact.

In the standard analysis, the primary input is the \SA for the primordial power spectrum, motivated by the assumption that the quantum state of cosmological perturbations is the BD vacuum at the onset of inflation. The basic finding of the LQC investigations is that the primordial spectrum of the \SA is in fact modified by pre-inflationary dynamics, but only for very small $k$, i.e., at very large angular scales, where it is no longer scale invariant.
This seems surprising at first because one expects quantum gravity effects to be significant at small (i.e. UV) scales rather than large (i.e. IR). However, there is an unforeseen UV-IR interplay that leads to this behavior. 

This origin of this interplay can be intuitively understood as follows \cite{aan}. The singularity resolution is indeed an UV effect --the matter density and curvature are rendered finite in the Planck regime because of UV corrections to Einstein's equations. Now, dynamics of the Fourier modes of cosmological perturbations is such that the time evolution of modes is affected by the presence of curvature only if their physical wavelength $\lambda_{\rm phy}$ is comparable to or greater than the curvature radius $\mathfrak{R}_{\rm curv} = \sqrt{6/R}$ where $R$ is the scalar curvature. This is just as one would expect physically:  if $\lambda_{\rm phy}  \ll \mathfrak{R}_{\rm curv}$,  the wave would propagate as though spacetime were flat (i.e. $\mathfrak{R}_{\rm curv} = \infty$). Now, while the scalar curvature $R$ diverges in GR,  it has a finite upper bound in LQC. Hence the curvature radius now has a non-zero lower bound, $\mathfrak{R}_{\rm min}$ \emph{which is a new scale provided by LQC.} Shorter wavelength modes with $\lambda \ll  \mathfrak{R}_{\rm min}$  do not experience spacetime curvature in their pre-inflationary evolution and for them pre-inflationary dynamics has negligible effect, and the primordial power spectrum is the same as that assumed in the SA. On the other hand, the long wavelength modes with $\lambda_{\rm phy} \gtrsim \mathfrak{R}_{\rm min}$ experience curvature in the Planck epoch near the bounce and get excited. These modes are not in the BD vacuum at the onset of the subsequent slow roll phase \cite{aan}. But would these excitations over the BD vacuum not just get diluted away during the 55+ e-folds of inflation? This was indeed a common expectation sometime ago. But the answer is in the negative  \cite{agullo-parker,ganc}: because of stimulated emission, the number \emph{density} of these excitations remains constant during inflation, whence the primordial spectrum at the end of inflation is modified. To summarize,  while singularity resolution is an UV phenomenon, the dynamics of cosmological perturbations is such that it is the \emph{longer} wavelength modes that receive significant LQC corrections in the primordial spectrum, breaking the near scale invariance assumed in the \SA for low wave numbers $k$. The precise departures from the \SA depend on the choice of the background quantum FLRW geometry $\Psi(a,\phi)$  and the quantum state  $\psi({\rm pert})$ of perturbations. 

\begin{figure}[]
  \begin{center}
  \includegraphics[width=3in,height=2.1in]{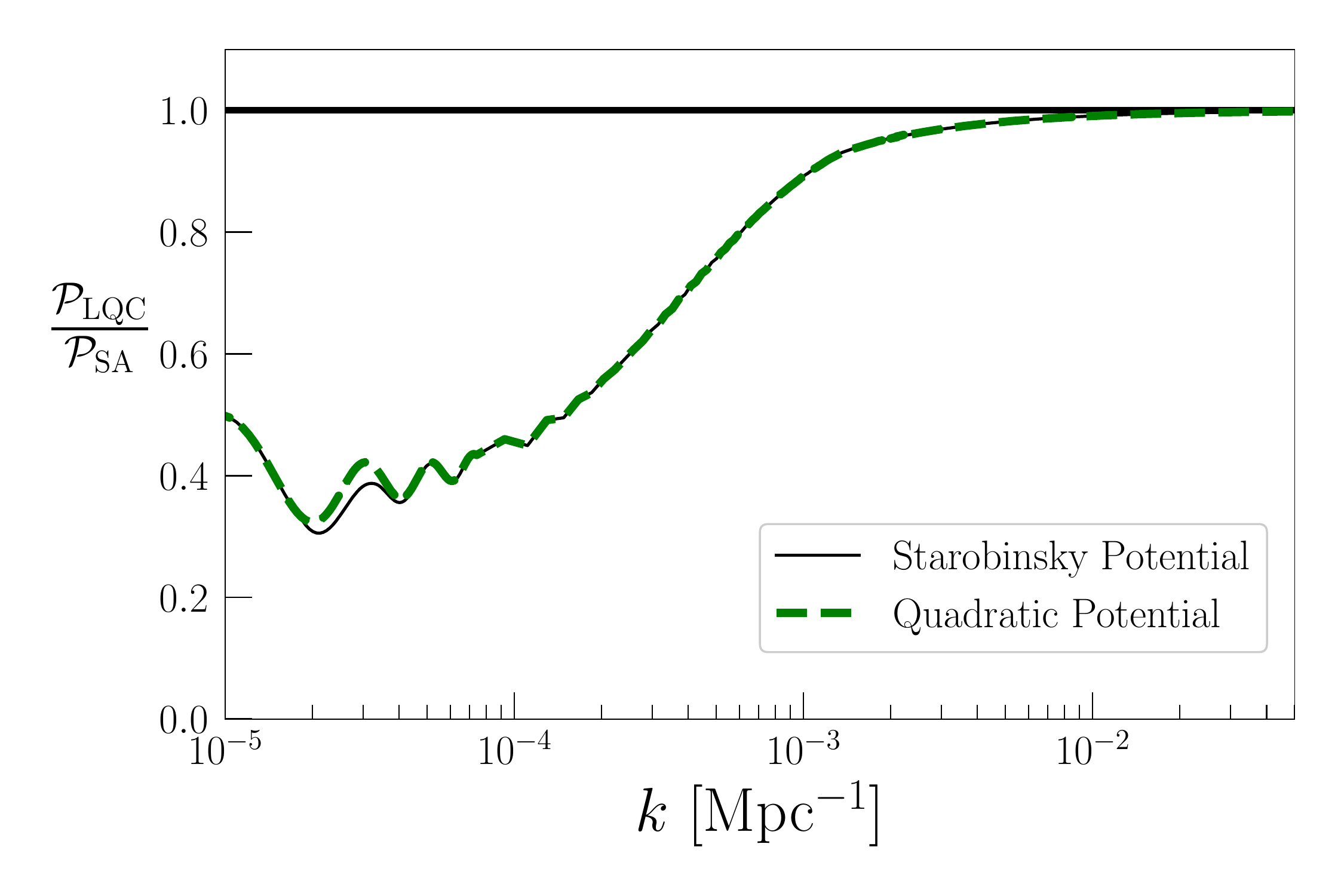}\hskip0.7cm
    \includegraphics[width=2.9in,height=2in]{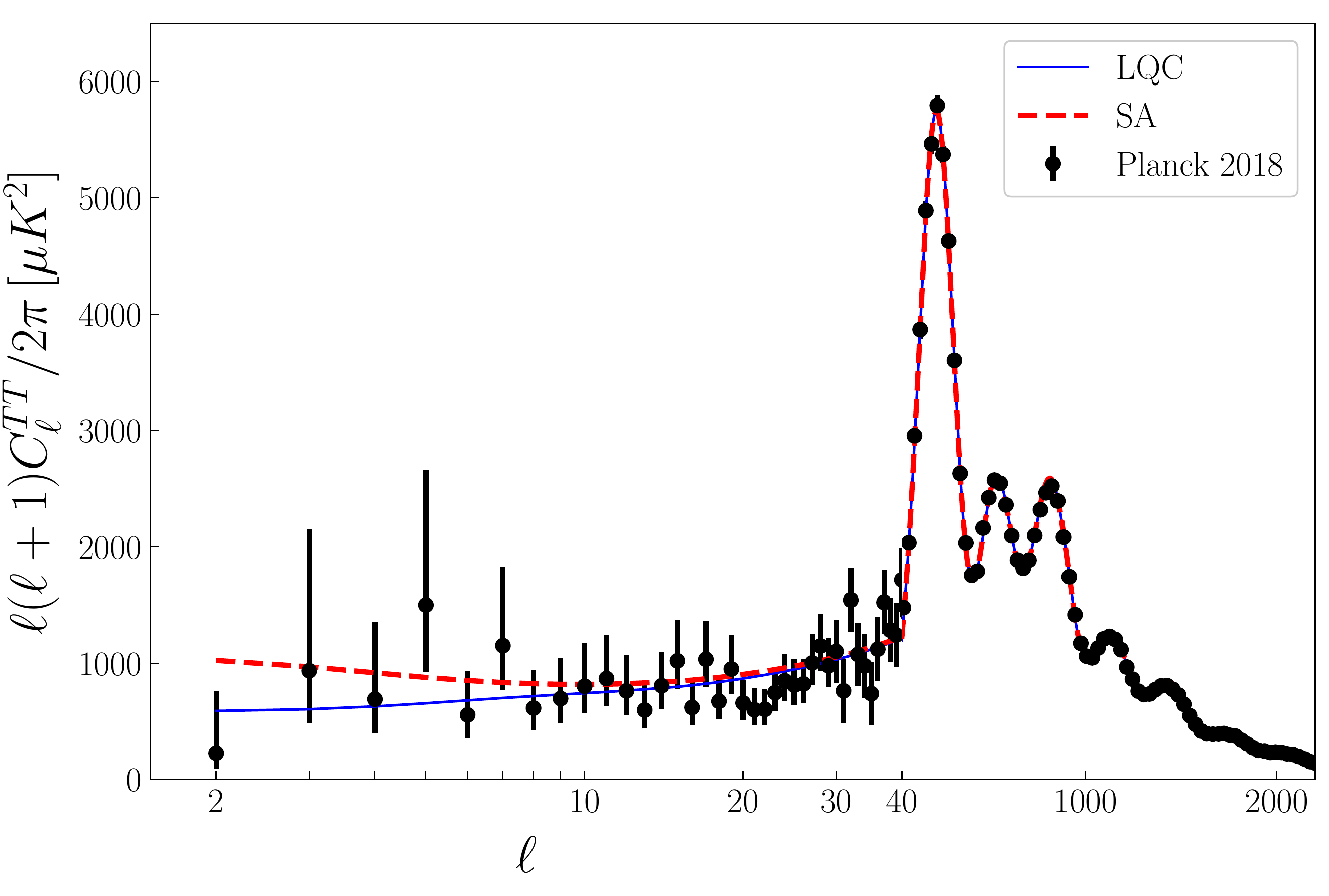}
\caption{LQC power suppression at large angular scales. \emph{Left Panel: Ratio of the primordial TT-power spectrum for LQC and SA.} Power is suppressed in LQC for $k\lesssim  3.6\times 10^{-3}{\rm Mpc}^{-1}$. Plots for the Starobinsky and quadratic potentials are essentially indistinguishable. \emph{Right Panel: Temperature-Temperature (TT) power spectra.} The 2018 CMB observations \cite{planck6} (black dots with error bars),  the LQC (solid (blue) line) and the standard ansatz  (SA) predictions (dashed (red) line). Credits:\cite{ag3,agjs}. }
 \label{fig:power}
\end{center}
\end{figure}

Several approaches have been pursued in LQC to investigate the observational consequences of the departure from the \SA (see, e.g.,  \cite{aan}-\cite{hybrid}). 
The key question is: Are modes with  $\lambda_{\rm phy} \gtrsim \mathfrak{R}_{\rm min}$ at the bounce in the range of wavelengths that CMB can observations can access today? The answer depends on the number of e-folds between the bounce and the onset of slow roll. If this number is too large, then these modes would not be accessible because the physical wavelength of these modes today would be larger than the radius of the visible universe. If on the other hand, the number is too small, then the LQC corrections would appear already at smaller wavelengths where the \SA works very well and LQC would be ruled out. For concreteness, to discuss this issue we will focus on a specific LQC approach \cite{ag3,agjs}. Here, the number of e-folds from the bounce to the onset of inflation is dictated by a general principle, inspired by quantum geometry at the bounce, and one finds that only modes with co-moving wave number $k \le 3.6 \times 10^{-3}\, {\rm Mpc}^{-1}$ receive significant LQC corrections in the primordial power spectrum. 
This corresponds to large angular scales, i.e., $\ell \lesssim 30$ in the $Y_{\ell, m}$ decomposition of the power spectrum. 

\begin{figure}[]
  \begin{center}
      \includegraphics[width=2.8in,height=2.2in]{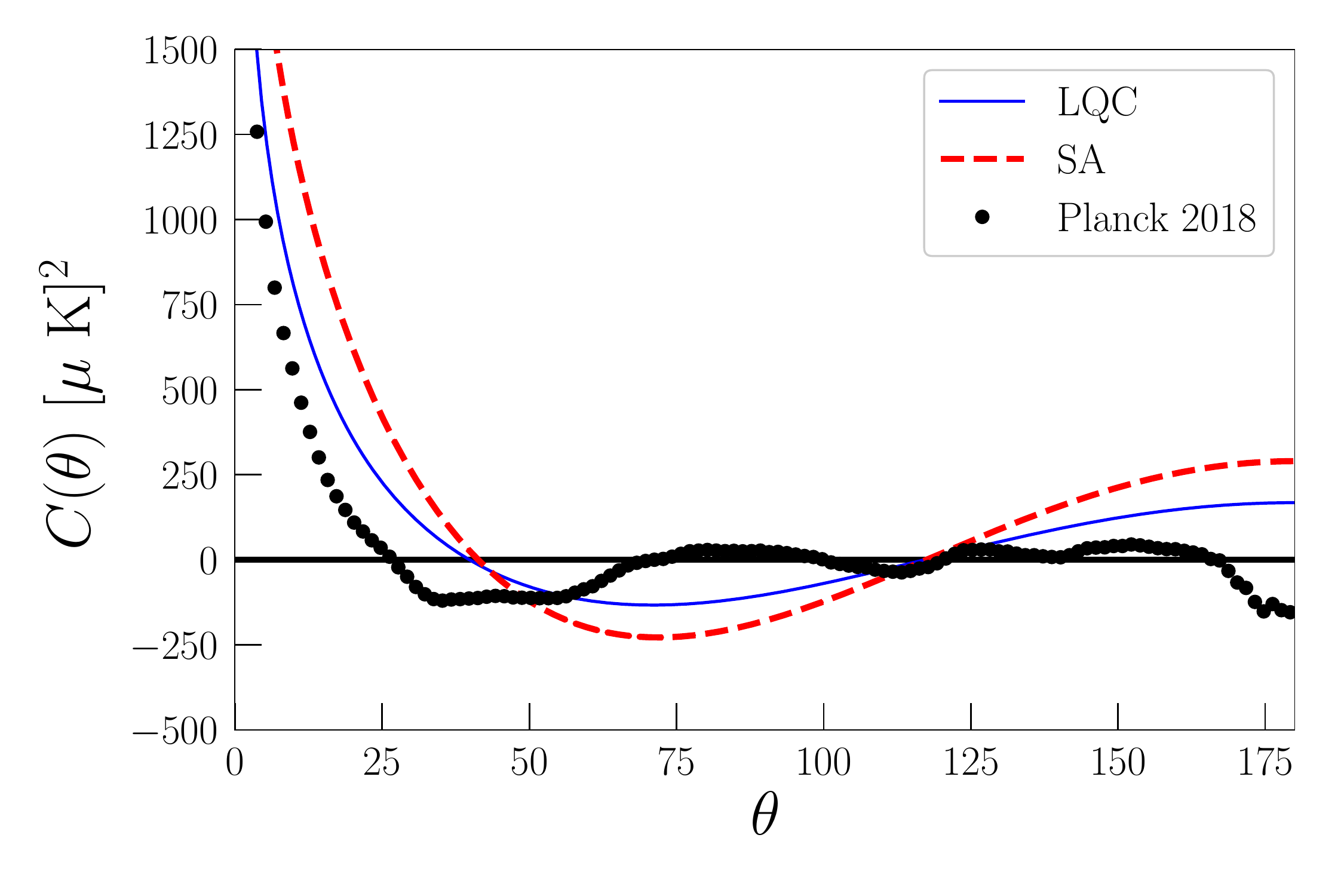}\hskip0.7cm
    \includegraphics[width=2.7in,height=2.1in]{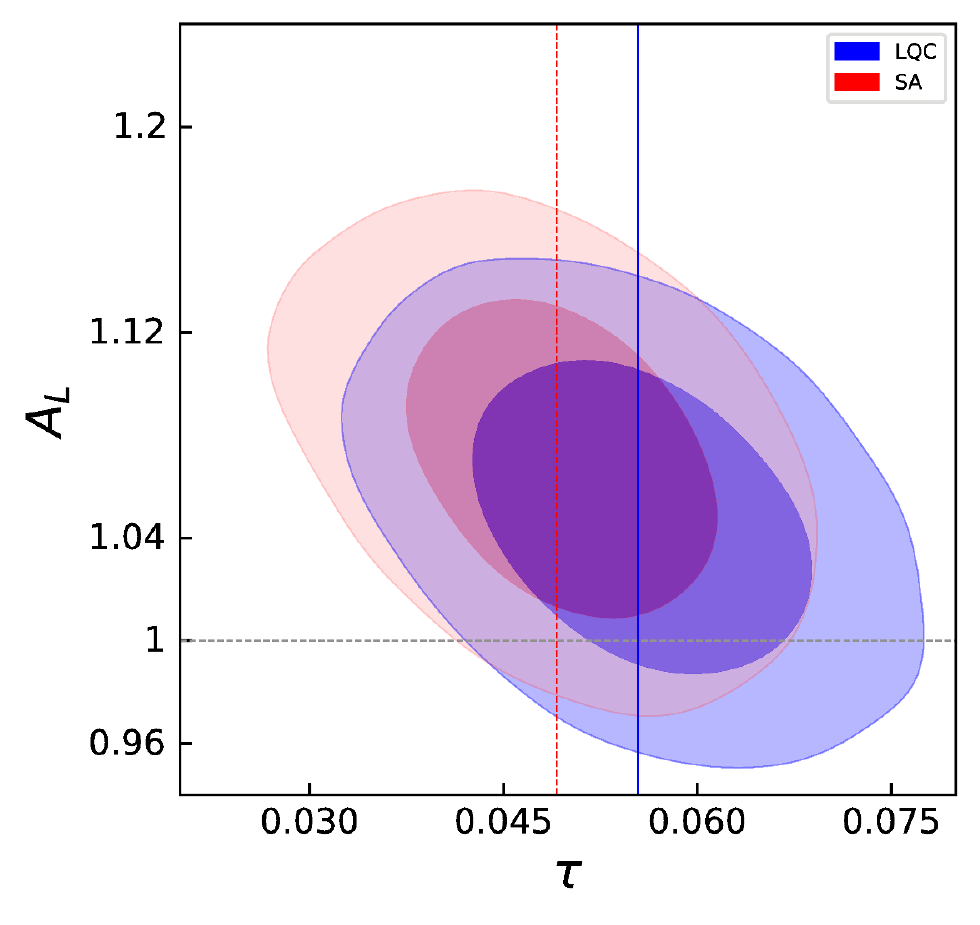}
    \caption{\emph{Left Panel:The angular power spectrum} $C(\theta) = \sum_{\ell}  (2\ell+1)\, C_{\ell} \, P_{\ell} (\cos\theta)$. The 2018 PLANCK-team spectrum (thick black dots),  the LQC  (solid (blue) line), and the standard ansatz (dashed (red) line) predictions.  \emph{Right Panel: $1\sigma$ and $2\sigma$ probability distributions in the $\tau\!-\!A_{L}$ plane.} Predictions of the standard ansatz (shown in red) and LQC (shown in blue). Vertical lines represent the respective mean values of $\tau$.}
\label{fig:ano}
\end{center}
\end{figure}

This modification of the primordial power spectrum then leads to the alleviation of two anomalies. 
The first is the so-called \emph{power suppression anomaly}: In the CMB, there is less power at $\ell \lesssim 30$ than that calculated using the \SA. 
The left panel of Fig.\ref{fig:power}  shows the status of the \emph{primordial} power spectrum in the LQC approach of \cite{ag3,agjs}. Already at the primordial level, there is a specific suppression relative to the \SA for low $k$, while the near scale invariance is maintained for large $k$.  As a consequence, as the right panel of Fig.\ref{fig:power} shows, the predicted T-T power spectrum is suppressed at large angular scales $\ell \le 30$ relative to the \SA and thus in better agreement with data.  As a result, had LQC+$\Lambda$CDM model been used for their analysis, the cosmic-variance uncertainties on large-scales would have been somewhat smaller than the values reported in \cite{planck6}! This power suppression at low $\ell$ was already observed by the WMAP team and several cosmologists have since argued that a better measure of power suppression is given by a quantity called $S_{1/2} := \int_{-1}^{1/2}\left[C(\theta)\right]^2 \rmd(\cos\theta)$, obtained by integrating the two-point T-T correlation function $C(\theta)$  over large angular scales ($\theta>60^\circ$) (see, e.g., \cite{shalf1,shalf2}). As the left panel of Fig. \ref{fig:ano} shows, the measured values of $S_{1/2}$ are significantly smaller than those predicted by the SA. The LQC prediction cuts this discrepancy by a factor of 3 \cite{agjs}. (Agreement with observations would be even better if the full LQC+$\Lambda$CDM model been used in the analysis of the PLANCK team.)  As a check on robustness, the LQC analysis was carried out using two different inflationary potential that have been widely used: the Starobinsky and quadratic potentials.

Because the LQC primordial power spectrum is different from that of standard inflation, the best fit values of the cosmological parameters are also different. Interestingly, while  5 of the 6 cosmological parameters are shifted by less than $0.4\%$, the LQC best fit value of the 6th --optical depth $\tau$-- is $9.8\%$ higher. Thus, the universe according to LQC is sufficiently different from that reported by the PLANCK collaboration \cite{planck6} (using the {\emph SA}) to have some observable consequences. One of these is the \emph{second anomaly, associated with the so-called lensing amplitude} $A_{L}$. Calculations leading to the cosmological model reported in \cite{planck6} require $A_{L}$=1. However, when it is allowed to vary,  $A_{L}$ prefers a value larger than unity, with $A_{L}$=1 lying \emph{outside} the 1-$\sigma$ contour. Recently it was suggested that this hint of internal inconsistency gives rise  to a ``possible crisis in cosmology'' \cite{silketal}.  However, thanks to the the higher value of the optical depth $\tau$ of LQC, the value $A_{L}$=1 is well within the  1-$\sigma$ contour (see the right panel of  Fig.\ref{fig:ano}). Thus, the LQC approach developed in \cite{ag3,agjs} resolves the tension associated with two anomalies simultaneously.  There is also an anomaly associated with hemispherical anisotropy: There exists a division of the celestial sphere into 2 hemispheres, with slightly higher average temperature in the southern 
hemisphere. This anomaly is explained in another investigation \cite{iaetal} within LQC 
that exploits the coupling between very long wavelength (super-horizon) modes with the observable ones. Thus, it is interesting that, thanks to the CMB observations, a possibility has opened up to see quantum gravity effects in the sky!

Interestingly, one can also go in the opposite direction and use observations to gain insight into quantum geometry. Recall that the area gap $\underline{\Delta}$ is the fundamental microscopic parameter in LQG.  Currently, its value is fixed by the requirement that the leading term in the statistical mechanical entropy of an isolated black hole, calculated using  the quantum horizon geometry, should yield the Bekenstein-Hawking formula (see, e.g., \cite{abck}-\cite{perez-review},\cite{alrev,30years:FBAP}). It is this value that is generally used in LQC. However, we can consider $\underline{\Delta}$ to be a free parameter in the CMB calculations, and find its posterior probability distribution by comparing the LQC theoretical predictions with observations. This would be an observational determination of $\underline{\Delta}$. It turns out that the value determined from the entropy considerations is well within 68\% confidence level of this distribution \cite{ag3}, i.e., within the standard error bars used by the PLANCK collaboration!

{\it Remark:} The power spectra, of course, depend on the choice of quantum states. In standard inflation, one makes this choice by demanding that the state be the BD vacuum at the start of the slow roll phase. In LQC it is natural to use the bounce time to specify the quantum states $\Psi(a,\phi)$ and $\psi({\rm pert})$. However, since during the pre-inflationary epoch the background space-time is not well-approximated by de Sitter geometry, one needs new inputs. Different LQC approaches use different strategies. In \cite{ag3,agjs}, for example, one uses certain principles to select these states. The overall strategy is to use such guidelines, work out the consequences, compare the predictions with observations, and abandon or refine the proposals depending on the success. The current success with a primordial explanation of anomalies signals that this program is viable. Can one perhaps combine the best features of different proposals to arrive at a compelling choice of the required quantum states? This is an outstanding open issue.

\section{Discussion}
\label{s5}

The last 4 sections  provide an impressionistic summary of the basic ideas and recent advances in LQG, addressed to beginning researchers in gravity and experts in other areas of physics. As explained in section \ref{s1}, because the  LQG literature is diverse and vast, we  chose to cover only a few topics to provide a coherent picture that illustrates the current status of the subject. In this section we will briefly discuss a few of the important advances that could not be included, and conclude by framing LQG in a broader context.

Quantum Riemannian geometry discussed in section \ref{s2} has been well established since the early 1990s. However, it continues to be enriched with new insights. For example,  while the salient features of quantum geometry emerged from a systematic investigation of representations of the holonomy-flux algebra $\Al$ \cite{lost}--\cite{almmt1}, recent investigations of gravity in presence of finite boundaries show that they arise much more generally from properties of the Casimir invariants associated with the boundary degrees of freedom \cite{boundary}.  Turning to dynamics, in section \ref{s3} we used spinfoams to illustrate the current status. As we saw, the program has tackled the difficult conceptual issues such as giving meaning to $n$-point functions in a diffeomorphism invariant context. It has also successfully met technical challenges of defining path integrals by an astute combination of quantum geometry, results from topological QFTs, and innovative numerical methods. The program has already been used to make contact with low energy physics via LQG calculations of the graviton propagator \cite{Bianchi:2009ri,Bianchi:2011hp}. However, as we indicated at the end of section \ref{s3.1}, important open issues remain at the foundational level, and considerable further work is still needed to bridge the gap between the Planck scale physics captured by the underlying quantum geometry of spinfoams and the rich effective field theory that has been developed over the years \cite{Donoghue:1994dn,Burgess:2003jk}.

There is a large number of results also in the Hamiltonian approach to LQG dynamics that we could not cover. Here, the aim to complete the program that Dirac \cite{dirac2} laid out  for a non-perturbative quantization of general relativity, emphasizing the fact that the dynamics is governed by constraints because of background independence. The well developed quantum kinematics, summarized in section \ref{s2}, provides the mathematical arena that is necessary for a precise definition of the constraint operators \cite{almmt1}. Through careful analysis of the underlying mathematical structures and introduction of astute and novel constructions, the `quantum spin dynamics' program led by Thiemann \cite{tt1} yielded well defined constraint operators needed in the Dirac program. Furthermore, in the matter sector, it provided concrete mechanisms for taming the UV divergences in the matter Hamiltonians through quantum geometry effects. However, this construction requires a number of choices in the gravitational part of the quantum Hamiltonian constraint, and the physical meaning and implications of these choices still remain unclear.  A more serious limitation is that, while these choices led to well-defined  quantum constraints that could be imposed without anomalies,  their algebra does not faithfully mirror the Poisson bracket algebra of classical constraints. More recently, this program was revived using new geometrical insights into the action of the Hamiltonian constraint at the classical level \cite{aamv}, provided by the gauge theory considerations of section \ref{s2.1}. It has been completed in several toy models that mimic the constraints of GR and also have an infinite number of degrees of freedom \cite{mv1}. Furthermore, these ideas have just been extended to full GR with Riemannian signature \cite{mv2}! The hope is that one can pass to the Lorentzian signature using the generalized Wick transform \cite{wick1}-\cite{wick3}. As we discuss below, there are already interesting investigations on the relation between full LQG and the symmetry reduced systems, such as LQC \cite{LQGLQC1,LQGLQC2}. Completion of the Hamiltonian dynamics program in full LQG will significantly sharpen these efforts and provide the much needed understanding of the underlying physics.  On the mathematical and conceptual side, advances to date in the Hamiltonian LQG dynamics are, by themselves, important. We chose not to discuss them in detail because, of necessity, they are quite technical in nature and  this review is addressed to non-experts. 

While  our understanding of quantum dynamics still remains rather far from being complete in full LQG,  advances could be made by using a \emph{truncation strategy}: One first chooses physical problems of interest --such as the early universe or black holes-- and focuses just on the sector of GR that is relevant to the problem. The program can be completed if the system admits a sufficient number of symmetries. Most notable progress in this direction has occurred in LQC, discussed in section \ref{s4.1}. In particular,  some of the key technical aspects of the recent constructions  \cite{mv1,mv2,30years:ALMV}  for dynamics of full LQG  were already mirrored in the Hamiltonian constraint of LQC \cite{aps}.  As we mentioned above, there has been considerable recent work on deriving the effective dynamics of LQC from full LQG  (see, e.g., \cite{LQGLQC1,LQGLQC2}).  Because LQG dynamics is still being developed, at this stage, certain choices have to be made to create this bridge. Therefore, currently different approaches lead to somewhat different results especially in the pre-bounce branch. However, in all these investigations, the big bang singularity is resolved and predictions for dynamics in the post big-bang branch are very similar to those of standard LQC, reported in section \ref{s4.1}.

A key non-trivial feature of the LQC dynamics is that corrections to GR are negligible until the matter density or curvature are $\sim$$10^{-4}$ of their Planck scale values but then they grow very rapidly, creating an effective repulsive force that completely overwhelm the classical attraction and causes the bounce.  Note that in this singularity resolution, matter does not violate the standard energy conditions. Yet the singularity theorems in classical GR are bypassed because the quantum corrections modify Einstein's equations themselves.  Perhaps the most striking feature of the developments in LQC is that, in addition to addressing the long-standing mathematical and conceptual issues of quantum cosmology, the subject has advanced sufficiently to make contact with observations. LQC  may well have predicted non-negligible quantum corrections on smaller angular scales, say $\ell \sim 100$, and therefore ruled out. That did not happen. Moreover, the LQC predictions alleviate the tension with the CMB anomalies \cite{ag3,agjs,iaetal}, providing a concrete realization of the possibility raised by the PLANCK collaboration \cite{planck1} that the anomalies may have a primordial origin (for the exact quote, see section \ref{s4.2}).  Furthermore, there is analysis of non-Gaussianities in one approach \cite{iaetal},  as well as predictions for the B-mode spectrum and a $\sim\!~ 10\% $ increase in the value of optical depth $\tau$  in another \cite{agjs}, which will be tested in upcoming observational missions. However, as we already noted, these are predictions of models within LQC; they are not direct consequences of full LQG. Nonetheless this bridge to observations reflects the extent to which the field has matured: through LQC, the program has left the pristine, high perch of mathematical physics to which most research in quantum gravity had been confined, and made a leap, joining other areas of physics such as QCD and gravitational waves where fundamental theory, model-building and experiments work in tandem.

There is one other important application of LQG  that we did not cover: quantum aspects of black holes.  As we briefly mentioned in section \ref{s4.2}, quantum Riemannian geometry has been used to provide a statistical mechanical account of the black hole entropy (see, e.g., \cite{abck}-\cite{perez-review},\cite{alrev,30years:FBAP}).
To begin with, the area gap $\underline{\Delta}$ is proportional to the undetermined Barbero-Immirzi parameter that arises in the passage from classical phase space to quantum kinematics. 
Now, in entropy calculations, one models a black hole in equilibrium by an isolated horizon and counts the surface degrees of freedom in the quantum horizon geometry that can interact with the exterior. A non-trivial result is that the logarithm of this number is proportional to the area. However,  the proportionality factor depends on $\underline{\Delta}$.  Now, it turns out that the geometry of an isolated horizon can be invariantly characterized via a set of multipoles. The strategy has been to use the simplest isolated horizon --the spherical one whose only non-zero multipole is the mass monopole--  to fix the value of the area gap $\underline{\Delta}$ such that the leading term in entropy for large black holes is given by the Bekenstein-Hawking formula. Once this is done, one can unambiguously calculate entropy of isolated horizons with arbitrary multipoles and verify that the leading term is again given by the Bekenstein-Hawking formula. Interestingly, as we noted in section \ref{s4.2}, CMB observations can be used to provide completely independent evidence in support of this value of the area gap! Next, there is now considerable ongoing  work to understand the evaporation process for non-rotating black holes, and the issue of information loss within LQG (for a recent review, see \cite{perez-review}).  There is growing evidence that quantum geometry would again lead to the resolution of the space-like singularity also in black holes, making the quantum spacetime vastly larger than in classical GR, just as in cosmology (see, e.g. \cite{eva1}-\cite{eva5}).  The basic premise is that the extended spacetime would admit complete future null infinity and the S-matrix from past null infinity to the future would be unitary. This possibility is being pursued in several directions within LQG. However, this is still work in progress and a number of open issues remain.

We will conclude by comparing and contrasting various approaches.  Let us begin with string theory.  As we already mentioned in section \ref{s1}, a number of bold ideas --such as supersymmetry, higher dimensions, extended objects and a negative cosmological constant-- constitute the foundations of string theory (see, e.g. Chapter 12 in \cite{centennial}) and were regarded as indispensable  for quantum gravity.  However, over the last 4 years or so, many of the leading practitioners have acknowledged that so far these ideas have failed to live up to their promise as a way to unite gravity and quantum mechanics,  and the focus of research has moved away from quantum gravity proper, to applications of techniques from GR  and supergravity to problems in an array of non-gravitational areas of physics.  As Dijkgraaf put it recently,  string theory researchers had been ``trying to aim for the successes of the past where we had a very simple equation that captured everything.''  ... But now  ``things have gotten almost postmodern'' \cite{ias}. String theory is now said to have embarked on a second life as one of the most useful sets of tools in science. 

By contrast, the focus in LQG continues to be on quintessentially quantum gravity issues. The underlying philosophy has been that quantum gravity should be rooted in well-established physics: principles of GR  and quantum mechanics. Ideas that have no observational support should not constitute an integral part of the foundation of quantum gravity, even when they can lead to rich mathematical structures. The starting point is just GR  coupled with matter. However, LQG is also radical in important ways. As we saw in section \ref{s2}, the fundamental quanta of geometry are very different from gravitons on flat spacetime. This balance between well-established principles and radical ideas is a hallmark of LQG.  Its key feature is the prominent role of the quantum nature of spacetime geometry.  In particular, as we saw, \emph{the area gap} --the lowest non zero eigenvalue of the area operator-- serves as  a basic microscopic parameter whose non-zero value leads to finite maximum values that matter density and curvature can attain in the standard cosmological models.  Thus, quantum geometry provides a natural, built-in ultraviolet cutoff.  In this respect it differs from other approaches such as Asymptotic Safety  or Dynamical Triangulations  that are more closely aligned with standard QFTs in the continuum (see, e.g., Chapter 11 in \cite{centennial} and \cite{loll2} ).  Because of its emphasis on quantum geometry and non-perturbative techniques,  as we saw in section \ref{s4}, LQG is well-placed to address the long standing problems of quantum gravity, such as the resolution of physically important singularities, the so-called `trans-Planckian issues' in the early universe cosmology, and the `problem of time'.  On the other hand, the very emphasis on quantum geometry and physics at the Planck scale has made it difficult for LQG to make rapid progress on establishing detailed contacts with low energy effective theories, or to uncover  implications of quantum gravity to the standard model of particle physics. By contrast, the Asymptotic Safety program, for example, has made significant progress in both these directions (see, e.g., \cite{eichhorn}).  Thus, because leading approaches to the problem of quantum gravity proper use diverse points of departure, reflecting the striking differences on where the primary emphasis should lie,  they have led to new insights in different directions, reflecting their complementary strengths. Given the difficulty of the task, this diversity is both healthy and essential.

\bigskip\bigskip
{\bf Acknowledgments:}  While writing this review we have drawn on illuminating scientific discussions we have had with almost everyone in  the LQG community over the years. In particular, the material covered in this review owes a great deal to Ivan Agullo, Pietro Don\`a, Brajesh Gupt, Jerzy Lewandowski, Carlo Rovelli, Parampreet Singh and Thomas Thiemann.  We would also like to thank Ed Wilson-Ewing for his comments on the first draft. This work was supported by NSF grants PHY-1806356,  PHY-1806428  and the Eberly endowment.


\bigskip\bigskip


\begin{thebibliography}{99}

\bibitem{einstein1916} Einstein, A  1916  N\"aherungsweise Integration der Feldgleichungen der Gravitation {\it Sitzungsberichte der Kniglich Preussischen Akademie der Wissenschaften (Berlin),} Seite 688-696 (the quote appears on p 696)

\bibitem{alrev} Ashtekar, A and Lewandowski, J  2004 {\it Background independent quantum gravity: A status report}, Class. Quant. Grav. {\bf 21} R53-R152 

\bibitem{crbook} Rovelli, C  2004 {\it Quantum Gravity}. (Cambridge: Cambridge UP)

\bibitem{crfvbook} Rovelli, C and Vidotto, F  2014 {\it Covariant Loop Quantum Gravity}  (Cambridge: Cambridge UP) 

\bibitem{ttbook} Thiemann,  T  2007 {\it Introduction to Modern Canonical Quantum General Relativity.}   (Cambridge: Cambridge UP) 

\bibitem{30years:KG} Giesel, K (2017) Quantum geometry, in \emph{Loop Quantum Gravity: The First 30 Years} ed. A. Ashtekar and J. Pullin (Singapore: World Scientific); Ch 1

\bibitem{30years:ALMV} Laddha, A and Varadarajan, M (2017) Quantum dynamics, Chapter 2 in the volume cited in \cite{30years:KG}

\bibitem{30years:IAPS} Agullo, I and Sigh, P (2017) Loop quantum cosmology, Chapter 6 in the volume cited in \cite{30years:KG}

\bibitem{30years:FBAP} Barbero, F and Perez A (2017) Quantum geometry and black holes,  Chapter 7 in the volume cited in \cite{30years:KG}

\bibitem{dirac2} Dirac, P A M (1964)  {\it Lectures on Quantum Mechanics, Belfer Graduate School Monograph Series, No.2}  (New York:Yeshiva University)

\bibitem{wheeler} Wheeler, J A (1964) in \emph{Relativity, Groups and Topology} ed. B S DeWitt and C M DeWitt (New York: Gordon Breach) 

\bibitem{ps1} Singh, P   2009  Are loop quantum cosmologies never singular? {\it Class. Quant. Grav.} \textbf{26} 125005

\bibitem{asrev} Ashtekar, A and Singh, P 2004  Loop quantum cosmology: A status report, 
{\it Class. Quant. Grav.} \textbf{28} 213001 (1-123)

\bibitem{centennial} Ashtekar, A  Berger, B  Isenberg, J and MacCallum M  (eds)   2015  \emph{General Relativity and Gravitation: A Centennial Perspective} (Cambridge U.P., Cambridge); Chapters 11 and 12

\bibitem{yang} Yang, C. N.  2005 \emph{Selected papers with commentary 1945-1980} (Singapore: World Scientific)

\bibitem{ias} Cole, K.C.  2016 The Strange Second Life of String Theory, Institute for Advanced Study News, https://www.ias.edu/news/cole-stringtheory-quanta

\bibitem{eichhorn} Eichhorn, A  2020  Asymptotically safe gravity,  \texttt{arXiv:2003.00044} [gr-qc]

\bibitem{bergmann} Bergmann, P G 1949 Non-linear field theories, {\it Phys. Rev.} \textbf{75}, 680-685 


\bibitem{dirac1} Dirac, P A M 1950 Generalized Hamiltonian Dynamics, {\it Can. J. Math.} \textbf{2}  129-148



\bibitem{adm} Arnowitt, R.  Deser, M. and Misner, C.W. 1962 The dynamics of general relativity, in \emph{Gravitation: Introduction to Current Research}  ed. L. Witten, (New York: Wiley), Ch 7

\bibitem{aamv} Ashtekar, A and Varadarajan, M 2021 Gravitational dynamics -A novel shift in the Hamiltonian paradigm {\it Universe} {\bf 7}13 (1-35)

\bibitem{aa-newvar} Ashtekar, A 1986  New variables for classical and quantum gravity,  {\it Phys. Rev. Lett.} \textbf{ 57}, 2244-2247\\
1987 A new Hamiltonian formulation of general relativity  {\it Phys. Rev.} D{\bf  36} 1587-1603 

\bibitem{aa-book} Ashtekar,  A 1991 \emph{Lectures on non-perturbative
canonical gravity}, (Notes prepared in collaboration with R.~S.~Tate)  (Singapore: World Scientific)

\bibitem{reula} Iriondo, M S  Leguiza\'{m}on, E O and Reula, O A (1997) Einstein's equations in Ashtekar's variables constitute a symmetric hyperbolic system  {\it Phys. Rev. Lett.} {\bf 79}, 4732-4735
 
\bibitem{shinkai} Yoneda, G and Shinkai,  H  1999  Symmetric hyperbolic system in the Ashtekar formulation  {\it Phys. Rev. Lett.} {\bf 82}  263-266 

\bibitem{art} Ashtekar, A  Romano, J and Tate, R S  1989 New variables for gravity: Inclusion of matter,  {\it Phys. Rev.} D{\bf 40} 2572-2587 


\bibitem{loops1} Rovelli C and Smolin L 1990 Loop representation for quantum
general relativity  \textit{Nucl. Phys.} \textbf{B331} 80--152
       
\bibitem{loops2} Gambini R and Pullin J 1996 \textit{Loops, knots, gauge
theories and quantum gravity}  (Cambridge UP, Cambridge)

\bibitem{wick1} Ashtekar, A 1996 A Generalized Wick transform for gravity {\it Phys.Rev.} D{\bf 53}:2865-2869

\bibitem{wick2} Thiemann, T   1996  Reality conditions inducing transforms for quantum gauge field theory and quantum gravity  {\it Class. Quant. Grav.} \textbf{13} 1383-1404

\bibitem{wick3} Varadarajan, M  2018 From Euclidean to Lorentzian loop quantum gravity via a positive complexifier  {\it Class. Quant. Grav.}  {\bf 36} 015016

\bibitem{fb} Barbero,  F  1996  Real Ashtekar variables for Lorentzian signature space-times {\it Phys. Rev.}  D{\bf 51} 5507-5510.

\bibitem{theta1}  Gambini,  R  Obregon, O and Pullin J  199 Yang-Mills analogues of the Immirzi ambiguity {\it Phys. Rev.}  D\textbf{59} 047505

\bibitem{theta2} Kaul R and 2012 Topological parameters in gravity {\it Phys. Rev.} D{\bf 85} 024026 


\bibitem{vonN} von Neumann J  1931 Die Eindeutigkeit der Schr\"odingerschen Operatoren, Math. Ann. \textbf{104}, 570-78; or in   1961 \emph{Collected Works,} Volume 2,  ed A. H. Taub (New York \& Oxford: Pergamon Press). 

\bibitem{hall} Hall, B C  2013  \emph{Quantum Theory for Mathematicians}, Graduate Texts in Mathematics, \textbf{267} (Berlin: Springer)

\bibitem{nonunique} Haag, R 1955  {\it Danske Vid. Selsk. Mat.-fys. Medd.} \textbf{29} No.12;\\
Garding, L  and Wightman,  A S  1956 {\it Proc. Nat. Acad. Sci. U.S.A.} \textbf{40} 622-626

\bibitem{segal} Segal, I E  1962 {\it Illinois J. Math.} {\bf 6} 500-523

\bibitem{lost} Lewandowski, L  Okolow, A   Sahlmann, H  and Thiemann, T 2006  Uniqueness of diffeomorphism invariant states on holonomy flux algebras, {\it Comm. Math. Phys.} \textbf{267} 703-733
  
\bibitem{cf} Fleishchack, C 2009 {Representations of the Weyl algebra in quantum geometry}, {\it Commun. Math. Phys.} \textbf{285} 67-140
 
 \bibitem{al2} Ashtekar  A and Lewandowski  J 1994 Representation theory of analytic holonomy algebras, in \textit{Knots and Quantum Gravity}  ed Baez J C (Oxford U.\ Press, Oxford)

\bibitem{almmt1} Ashtekar  A, Lewandowski  J, Marolf  D, Mour\~ao  J
and Thiemann  T 1995 Quantization of diffeomorphism invariant theories of connections with local degrees of freedom  {\it Jour. Math. Phys.} \textbf{36} 6456--6493

\bibitem{rp} Penrose  R 1971 Angular momentum: an approach to combinatorial space-time  \textit{Quantum Theory and Beyond}  ed T Bastin (Cambridge: Cambridge University Press)

\bibitem{loll1} Loll, R 1995  The volume operator in discretized quantum gravity  {\it Phys. Rev. Lett.} \textbf{75} 3048 

\bibitem{rs4} Rovelli C and Smolin  L 1995  Discreteness of area and volume in quantum gravity  {\it Nucl. Phys.} B{\bf 442}  593--622; Erratum: \textit{Nucl. Phys.} B {\bf 456}  753

\bibitem{al4} Ashtekar A and Lewandowski  L 1995  Differential geometry on the space of connections using projective techniques {\it Jour. Geo. \& \ Phys.} \textbf{17}  191--230

\bibitem{al5} Ashtekar  A and Lewandowski  J 1997  Quantum theory of geometry I: Area operators  {\it Class. Quant. Grav.} \textbf{14}  A55--A81

\bibitem{al6} Ashtekar  A and Lewandowski  J 1997  Quantum theory of geometry II: Volume Operators  {\it Adv. Theo. Math. Phys.} \textbf{1}  388--429

\bibitem{tt2} Thiemann  T 1998  A length operator for canonical quantum gravity  {\it Jour. Math. Phys.} {\bf 39}  3372--3392

\bibitem{eb1} Bianchi, E  2009  The Length Operator in Loop Quantum Gravity  {\it Nucl. Phys.} B\textbf{807} 591-624

\bibitem{hstt} Sahlmann, H and Thiemann, T 2006\\
Towards the QFT on curved space-time limit of QGR. 1. A General scheme {\it Class. Quant. Grav.} \textbf{23}  867-908\\
Towards the QFT on curved space-time limit of QGR.  2. A Concrete implementation  {]\it Class. Quant .Grav.} \textbf{23} 909-954


\bibitem{Misner:1957wq}
Misner, C W 1957 {Feynman quantization of general relativity},'' {\em Rev. Mod.  Phys.} {\bf 29}  497--509

\bibitem{Hawking:1978jz}
Hawking, S W 1978 {Quantum Gravity and Path Integrals}, {\em Phys. Rev.} {\bf  D18} 1747--1753

\bibitem{Reisenberger:1996pu}
Reisenberger, M P and Rovelli, C 1997 {'Sum over surfaces' form of loop quantum gravity}, {\em Phys. Rev.} {\bf D56} 3490--3508

\bibitem{Perez:2012wv}
Perez, A 2013 {The Spin Foam Approach to Quantum Gravity}, {\em Living Rev. Rel.} {\bf 16} 3

\bibitem{30years:EB} Bianchi, E (2017) Spinfoam gravity, Chapter 3 in the volume cited in \cite{30years:KG}

 \bibitem{Baez:1999sr}
Baez, J C 2000 {An Introduction to spin foam models of quantum gravity and BF
  theory}, {\em Lect. Notes Phys.} {\bf 543} 25--94
  
\bibitem{Horowitz:1989ng}
Horowitz, G T 1989 {Exactly Soluble Diffeomorphism Invariant Theories}, {\em Commun. Math. Phys.} {\bf 125} 417

\bibitem{Witten:1988ze}
Witten, E 1988 {Topological Quantum Field Theory}, {\em Commun. Math. Phys.} {\bf 117} 353

\bibitem{Atiyah:1989vu}
Atiyah, M 1989 {Topological Quantum Field Theories}, {\em Inst. Hautes Etudes Sci. Publ. Math.} {\bf 68} 175--186

\bibitem{Holst:1995pc}   
Holst, S 1996 Barbero's Hamiltonian derived from a generalized Hilbert-Palatini action \textit{Phys. Rev.} \textbf{D53} 5966-5969

\bibitem{Wieland:2010ec}
Wieland, W 2012 Complex Ashtekar variables and reality conditions for Holst's action, Annales Henri Poincare \textbf{13}, 425-448

\bibitem{Baez:1994zz}
Baez, J 1994 {Knots and quantum gravity: Progress and prospects}, Proceedings of the Seventh Marcel Grossman Meeting, arXiv:gr-qc/9410018 [gr-qc]

\bibitem{Plebanski:1977zz}
Plebanski, J F 1977 {On the separation of Einsteinian substructures}, {\em J. Math. Phys.} {\bf 18} 2511--2520

\bibitem{Capovilla:1989ac}
Capovilla, R, Jacobson, T, and Dell, J 1989 {General Relativity Without the Metric}, {\em Phys. Rev. Lett.} {\bf 63} 2325

\bibitem{Reisenberger:1996ib}
Reisenberger, M P 1997 {A Left-handed simplicial action for Euclidean general relativity}, {\em Class. Quant. Grav.} {\bf 14} 1753--1770

\bibitem{Bianchi:2009tj}
Bianchi, E 2014 {Loop Quantum Gravity \`a la Aharonov-Bohm}, {\em Gen. Relativ. Gravit.} {\bf 46} 1668

\bibitem{Regge:1961px}
Regge, T 1961 {General relativity without coordinates}, {\em Nuovo Cim.} {\bf 19} 558--571

\bibitem{Oeckl:2005rh}
Oeckl, R 2005 {Discrete gauge theory: From lattices to topological quantum field theory}, (Cambridge: Cambridge UP)

\bibitem{Engle:2007qf}
Engle, J, Pereira, R, and Rovelli, C, 2008 {Flipped spinfoam vertex and loop gravity}, {\em Nucl. Phys.} {\bf B798} 251--290

\bibitem{Freidel:2007py}
Freidel, L and Krasnov, K 2008 {A New Spin Foam Model for 4d Gravity}, {\em Class. Quant. Grav.} {\bf 25} 125018
  
\bibitem{Engle:2007wy}
Engle, J, Livine, E, Pereira, R and Rovelli, C 2008 {LQG vertex with finite Immirzi parameter}, {\em Nucl. Phys.} {\bf B799} 136--149

\bibitem{Kaminski:2009fm}
Kaminski, W, Kisielowski, M and Lewandowski, J 2010 {Spin-Foams for All Loop Quantum Gravity}, {\em Class. Quant. Grav.} {\bf 27} 095006 [Erratum: Class. Quant. Grav.29,049502(2012)]


\bibitem{Kaminski:2011bf}
Kaminski, W, Kisielowski, M and Lewandowski, J 2012
{The Kernel and the injectivity of the EPRL map},
{\em Class. Quant. Grav.} \textbf{29}, 085001

\bibitem{Ding:2010fw}
Ding, Y, Han, M and Rovelli, C 2011
{Generalized Spinfoams},
{\em Phys. Rev. D}\textbf{83}, 124020

\bibitem{Thiemann:2013lka}
Thiemann, T and Zipfel, A, 2014
{Linking covariant and canonical LQG II: Spin foam projector},
{\em Class. Quant. Grav.} \textbf{31}, 125008

\bibitem{Martin-Dussaud:2019ypf}
Martin-Dussaud, P 2019
A Primer of Group Theory for Loop Quantum Gravity and Spin-foams,
Gen. Relativ. Gravit. \textbf{51}, no.9, 110 

\bibitem{Ponzano}
Ponzano G and Regge T 1968 Semiclassical limit of Racah coeffecients, {\em Spectroscopic and Group Theoretical Methods in Physics, edited by F. Block (North Holland, Amsterdam)}

\bibitem{Han:2010pz}
Han, M 2011 {4-dimensional Spin-foam Model with Quantum Lorentz Group}, {\em J.
  Math. Phys.} {\bf 52} 072501
  
\bibitem{Fairbairn:2010cp}
Fairbairn, W J and Meusburger, C 2012 {Quantum deformation of two
  four-dimensional spin foam models}, {\em J. Math. Phys.} {\bf 53}
  022501

\bibitem{Haggard:2014xoa}
Haggard, H M, Han, M, Kaminski, W and Riello, A 2015 {SL(2,C) Chern-Simons Theory, a non-Planar Graph Operator, and 4D Loop Quantum Gravity with a Cosmological Constant: Semiclassical Geometry}, {\em Nucl. Phys.} {\bf B900} 1--79


\bibitem{Riello:2013bzw}
Riello, A 2013
{Self-energy of the Lorentzian Engle-Pereira-Rovelli-Livine and Freidel-Krasnov model of quantum gravity,}
{\em Phys. Rev. D}\textbf{88}, no.2, 02401

\bibitem{Dona:2018pxq}
Don\`a, P 2018
{Infrared divergences in the EPRL-FK Spin Foam model},
{\em Class. Quant. Grav.} \textbf{35}, no.17, 175019

\bibitem{Rovelli:2010qx}
Rovelli, C and Smerlak, M 2012,
{In quantum gravity, summing is refining,}
{\em Class. Quant. Grav.} \textbf{29}, 055004



\bibitem{DePietri:1999bx}
De Pietri, R Freidel, L Krasnov, K and Rovelli, C 2000
Barrett-Crane model from a Boulatov-Ooguri field theory over a homogeneous space,
Nucl. Phys. B \textbf{574}, 785-806

\bibitem{Oriti:2013aqa}
Oriti, D 2016 Group field theory as the 2nd quantization of Loop Quantum Gravity,
Class. Quant. Grav. \textbf{33}, no.8, 085005 

\bibitem{Donoghue:1994dn} 
Donoghue, J F 1994
General relativity as an effective field theory: The leading quantum corrections, Phys. Rev. D \textbf{50}, 3874-3888

\bibitem{Burgess:2003jk}
Burgess, C P 2004 Quantum gravity in everyday life: General relativity as an effective field theory, Living Rev. Rel. \textbf{7}, 5-56

\bibitem{Rovelli:2005yj}
Rovelli, C 2006 {Graviton propagator from background-independent quantum gravity}, {\em Phys. Rev. Lett.} {\bf 97} 151301

\bibitem{Bianchi:2006uf}
Bianchi, E, Modesto, L, Rovelli, C and Speziale, S 2006 {Graviton propagator in loop quantum gravity}, {\em Class. Quant. Grav.} {\bf 23}
  6989--7028

\bibitem{Thiemann:2000bw}
Thiemann, T 2001 {Gauge field theory coherent states (GCS): 1. General
  properties}, {\em Class. Quant. Grav.} {\bf 18} (2001) 2025--2064
 
\bibitem{Bahr:2007xn}  
Bahr, B and Thiemann, T 2009 {Gauge-invariant coherent states for loop quantum  gravity. II. Non-Abelian gauge groups}, {\em Class. Quant. Grav.} {\bf 26} 045012

\bibitem{Bianchi:2009ky}
Bianchi, E, Magliaro, E and Perini, C 2010 {Coherent spin-networks}, {\em Phys. Rev.} {\bf D82} 024012

\bibitem{Freidel:2010tt}
Freidel, L and Livine, E R 2011 U(N) Coherent States for Loop Quantum Gravity,
J. Math. Phys. \textbf{52}, 052502

\bibitem{Calcinari:2020bft}
Calcinari, A, Freidel, L, Livine, E and Speziale, S 2020
{Twisted Geometries Coherent States for Loop Quantum Gravity,}
{\em Class. Quant. Grav.} \textbf{38}, no.2, 025004

\bibitem{Bianchi:2016hmk}
Bianchi, E, Guglielmon, J, Hackl, L and Yokomizo, N 2016
{Loop expansion and the bosonic representation of loop quantum gravity},
{\em Phys. Rev. D}\textbf{94}, no.8, 086009

\bibitem{Bianchi:2010gc}
Bianchi, E, Dona, P and Speziale, S 2011 {Polyhedra in loop quantum gravity},  {\em Phys. Rev.} {\bf D83} (2011) 044035

\bibitem{Bianchi:2011ub}
Bianchi, E and Haggard, H M 2011 {Discreteness of the volume of space from Bohr-Sommerfeld quantization}, {\em Phys. Rev. Lett.} {\bf 107}
  011301

\bibitem{Dittrich:2008ar}
Dittrich, B and Ryan, J P 2011 {Phase space descriptions for simplicial 4d geometries}, {\em Class. Quant. Grav.} {\bf 28} 065006

\bibitem{Freidel:2010aq}
Freidel, L and Speziale, S 2010 {Twisted geometries: A geometric parametrisation of SU(2) phase space}, {\em Phys. Rev.} {\bf D82} 2010 084040 

\bibitem{Bianchi:2012ev}
Bianchi, E and Myers, R C
2014 On the Architecture of Spacetime Geometry,''
Class. Quant. Grav. \textbf{31}, 214002

\bibitem{Baytas:2018wjd}
Baytas, B  Bianchi, E and Yokomizo, N 2018
Gluing polyhedra with entanglement in loop quantum gravity,
Phys. Rev. D \textbf{98}, no.2, 026001

\bibitem{Oeckl:2005bv}
Oeckl, R  2008 {General boundary quantum field theory: Foundations and probability
  interpretation}, {\em Adv. Theor. Math. Phys.} {\bf 12} 319--352
  
\bibitem{Barrett:2009gg}
Barrett, J W, Dowdall, R J, Fairbairn, W J, Gomes, H and Hellmann, F 2009
  {Asymptotic analysis of the EPRL four-simplex amplitude}, {\em J. Math.  Phys.} {\bf 50} 112504
  
\bibitem{Barrett:2009mw}
Barrett, J W, Dowdall, R J, Fairbairn, W J, Hellmann, F and Pereira, R 2010
  {Lorentzian spin foam amplitudes: Graphical calculus and asymptotics}, {\em Class. Quant. Grav.} {\bf 27} 165009

\bibitem{Dona:2019dkf}
Don\`a, P, Fanizza, M, Sarno, G and Speziale, S 2019
Numerical study of the Lorentzian Engle-Pereira-Rovelli-Livine spin foam amplitude, Phys. Rev. D \textbf{100}, no.10, 106003

\bibitem{Bianchi:2009ri} 
Bianchi, E, Magliaro, E and Perini, C 2009 {LQG propagator from the new spin foams}, {\em Nucl. Phys.} {\bf B822} 245--269

\bibitem{Bianchi:2011hp}
Bianchi, E and Ding, Y 2012 {Lorentzian spinfoam propagator}, {\em Phys. Rev.}
  {\bf D86} 104040

\bibitem{Bonzom:2009hw}
Bonzom, V 2009
{Spin foam models for quantum gravity from lattice path integrals},
{\em Phys. Rev.} D\textbf{80}, 064028 (2009)

\bibitem{Engle:2020ffj}
Engle, J S, Kaminski, W and Oliveira, J R 2020
{Comment on ``EPRL/FK Asymptotics and the Flatness Problem''},
arXiv:2012.14822 [gr-qc]

\bibitem{Dona:2020tvv}
Dona, P Gozzini, F and Sarno, G 2020
Numerical analysis of spin foam dynamics and the flatness problem,
{\em Phys. Rev. D}\textbf{102}, no.10, 106003



\bibitem{Asante:2020qpa}
Asante, S K, Dittrich, B and Haggard, H M 2020
Effective Spin Foam Models for Four-Dimensional Quantum Gravity,
Phys. Rev. Lett. \textbf{125}, no.23, 231301

\bibitem{Asante:2020iwm}
Asante, S K, Dittrich, B and Haggard, H M 2020
{Discrete gravity dynamics from effective spin foams},
arXiv:2011.14468 [gr-qc]

\bibitem{Bahr:2015gxa}
Bahr, B and Steinhaus, S 2016 Investigation of the Spinfoam Path integral with Quantum Cuboid Intertwiners,
{\em Phys. Rev. D}\textbf{93}, no.10, 104029

\bibitem{Gozzini:2019kui}
Dona, P Gozzini, F and Sarno, G 2020
{Searching for classical geometries in spin foam amplitudes: a numerical method},
{\em Class. Quant. Grav.} \textbf{37}, no.9, 094002

\bibitem{Gozzini:2019nbo}
Gozzini, F and Vidotto, F 2019 Primordial fluctuations from quantum gravity,
arXiv:1906.02211

\bibitem{Han:2020fil}
Han, M, Huang, Z, Liu, H and Qu, D 2020
{Numerical computations of next-to-leading order corrections in spinfoam large-$j$ asymptotics},
{\em Phys. Rev. D}\textbf{102}, no.12, 124010

\bibitem{Han:2020npv}
Han, M, Huang, Z, Liu, H, Qu, D and Wan, Y 2020
{Spinfoam on Lefschetz Thimble: Markov Chain Monte-Carlo Computation of Lorentzian Spinfoam Propagator},
arXiv:2012.11515 [gr-qc]



\bibitem{einstein1946} Einstein A 1946 \emph{The Meaning of Relativity}, Third Edition (Princeton: Princeton UP), App. I 

\bibitem{halper} Ashtekar, A  Efstathiou, G  Guth, A  Hawking, S W  Penrose, R  Veneziano, G and Vilenkin, A   2019 \emph{The Big Bang's new meaning,}  \href{https://www.youtube.com/watch?v=U7kvjTRW-tw}{www.youtube.com/watch?v=U7kvjTRW-tw} 

\bibitem{mb} Bojowald  M 2001  Absence of singularity in loop quantum cosmology  \textit{Phys. Rev. Lett.} \textbf{86} 5227--5230

\bibitem{aps} Ashtekar, A  Pawlowski, T and Singh P  2006 {Quantum nature of the big bang}, {\it Phys. Rev. Lett.} \textbf{96} 141301\\
 {Quantum nature of the big bang: Improved dynamics}, {\it Phys. Rev.} D{\bf 74} 084003

\bibitem{cs1} Corichi, A and Singh, P 2008 Is loop quantization in cosmology unique? {\it Phys. Rev.}  D\textbf{78} 024034 

\bibitem{acs}  Ashtekar, A Corichi, A and Singh, P 2008  Robustness of key features of loop quantum cosmology {\it Phys. Rev.} D\textbf{77}  024046

\bibitem{gowdy} Navascu\'es, B E Martin-Benito, M and Mena Marug\'an G (2015)  Modified FRW cosmologies arising from states of the hybrid quantum Gowdy model {\it Phys. Rev.} D\textbf{92} 024007
 
\bibitem{bdtheory} Zhang, X  Ma, Y and Artymowski, M 2013 Loop quantum Brans-Dicke cosmology {\it Phys. Rev.} D\textbf{8} 084024

\bibitem{mb2} Bojowald, M 2020 Critical evaluation of common claims in loop quantum cosmology,  \textit{Universe} \textbf{6}  36

\bibitem{aa-grg} Ashtekar, A 2009 Loop quantum cosmology: An overview  {\it Gen. Relativ. Gravit.} {\bf 41} 707-741  

\bibitem{kp} Kaminski, W  and Pawlowski, T 2010,  Cosmic recall and the scattering picture of Loop Quantum Cosmology  {\it Phys. Rev.} D{\bf 81} 084027 

\bibitem{ac} Ashtekar, A and Campiglia M 2012   On the uniqueness of kinematics of loop quantum cosmology,  {\it Class. Quant. Grav.}  \textbf{29}, 242001 
    
 \bibitem{eht} Engle, J  Hanusch, M and Thiemann, T 2017  Uniqueness of the representation in homogeneous isotropic LQC  {\it Commun. Math. Phys.} \textbf{354}, 231-246 

\bibitem{planck1} Aghanim, N et al 2020  Planck2018 results  I. Overview and the cosmological legacy of Planck,  {\it Astron. \& Astrrophys.} \textbf{641} A1  

\bibitem{planck6} R.~Adam {\it et al.} [Planck Collaboration], Planck 2018 results VI. Cosmological parameters, {\it Astronomy \& Astrophysics} \textbf{641} A6 (2020).

\bibitem{aan} Agullo, I  Ashtekar, A  and Nelson W 2012 A Quantum Gravity Extension of the Inflationary Scenario, {\it Phys. Rev. Lett.}  {\bf 109}  251301\\
2013 Pre-inflationary dynamics of loop quantum cosmology: Confronting quantum gravity with observations, {\it Class. Quant. Grav.}  {\bf 30} 085014 1-56 

\bibitem{agullo-parker} Agullo, I  and Parker, L 2011  Non-gaussianities and the stimulated
 creation of quanta in the inflationary universe, {\it Phys. Rev.} D{\bf 83} 063526

\bibitem{ganc} Ganc,  J. 2011 Calculating the local-type fNL for slow-roll inflation with a non-vacuum initial state., {\it Phys. Rev.}  D{\bf 84} 063514 

\bibitem{barrauetal} Barrau, A  Cailleteau, T  Grain, J  and Mielczarek,  J  2014 Observational issues in loop quantum cosmology,  {\it Class. Quant. Grav.} {\bf 31}  053001,

\bibitem{ewe} Wilson-Ewing, E  2017 Testing loop quantum cosmology, {\it Compt. Ren. Phys.} {\bf 18},  207-225 

\bibitem{ag3}  Ashtekar, A and  Gupt, B 2017 Quantum gravity in the sky: Interplay between fundamental theory and observations, {\it Class. Quant. Grav.}   {\bf 34}, 014002 

\bibitem{agjs} Ashtekar, A  Gupt B  Jeong, D and Sreenath V 2020  Alleviating the tension in CMB using Planck-scale Physics {\it Phys. Rev. Lett.} \textbf{125} 051302
 
 \bibitem{iaetal} Agullo, I  Kranas, D and Sreenath V 2020 Large scale anomalies in the CMB and non-Gaussianity in bouncing cosmologies, arXiv:2006.09605
 
 \bibitem{hybrid} Navascu\'es, B E  and Mena-Marug\'an G 2020  Hybrid loop quantum cosmology: An overview  arXiv:2011.04559
 
 \bibitem{shalf1}  Sarkar, D  Huterer, D  Copi, C J   Starkman G D and Schwarz, D J (2011) 
 Missing power vs low-l alignments in the cosmic microwave background: No correlation in the standard cosmological model  {\it Astropart. Phys}  {\bf 34}, 591 

\bibitem{shalf2} Schwarz, D J  Copi, C J   Huterer, D and Starkman G D  2016 CMB anomalies after Planck  {\it Class. Quant. Grav.}  {\bf 33} 184001 

\bibitem{silketal}  Di Valentino, E  Melchiorri, A and Silk, J 2020 Planck evidence for a closed Universe and a possible crisis for cosmology {\it Nat.  Astron.} 196-203


\bibitem{boundary} 
Freidel, L Geiller, M and Pranzetti, D 2020 \\
Edge modes of gravity I: Corner potentials and charges, JHEP \textbf{11}, 026\\
Edge modes of gravity II: Corner metric and Lorentz charges, JHEP \textbf{11}, 027\\
Edge modes of gravity III: Corner simplicity constraints, \texttt{arXiv:2007.12635}

\bibitem{tt1} Thiemann, T  1998\\  
Quantum spin dynamics (QSD)  {\it Class. Quant. Grav.} \textbf{15} 839-873\\
Quantum spin dynamics (QSD) II  {\it Class. Quant. Grav.} 875-905\\ 
Quantum spin dynamics (QSD) III: Quantum constraint algebra and physical scalar product in quantum general relativity   {\it Class. Quant. Grav.} 1207-1247\\
Quantum spin dynamics (QSD) IV: 2+1 Euclidean Quantum Gravity as a model to test 3+1 Lorentzian Quantum Gravity  {\it Class. Quant. Grav.} 1249-1280 \\
Quantum spin dynamics (QSD) V: Quantum gravity as the natural regulator of matter quantum field theories 1281-1314 

\bibitem{mv1} Varadarajan, M 
2018  The constraint algebra in Smolin's $G \to 0$ limit of 4d Euclidean Gravity {\it Phys. Rev.} D{\bf 97} 106007\\
2019  On quantum propagation in Smolin's weak coupling limit of 4d Euclidean Gravity  
{\it Phys.Rev.} D{\bf 100}  066018

\bibitem{mv2} Varadarajan, M 2020 Euclidean LQG dynamics: An electric shift in perspective, arXiv: 2101.03115

\bibitem{LQGLQC1} Assanioussi, M  Dapor, A  Liegener, K  Pawlowski, T 2018  Emergent de Sitter epoch of the quantum cosmos from loop quantum cosmology  {\it Phys. Rev. Lett.} {\bf 121} 081303

\bibitem{LQGLQC2} Olmedo, J and Alesci, E  2019  Power spectrum of primordial perturbations for an emergent universe in quantum reduced loop gravity,  {\it JCAP} {\bf 04} 030

\bibitem{loll2} Loll, R  2019 Quantum gravity from causal dynamical triangulations: a review, Class. Quantum Grav. \textbf{37} 013002

\bibitem{abck} Ashtekar, A  Baez, J  Corichi A and Krasnov K 1998 Quantum geometry and black hole entropy  (1998) {\it Phys. Rev. Lett.} \textbf{80} 904--907\\
Ashtekar  A, Baez  J C   and Krasnov  K 2000 Quantum geometry of isolated horizons and black hole entropy  \textit{Adv. Theo. Math. Phys.}  \textbf{4}  1--95

\bibitem{enpp} Engle, J  Noui K  Perez A and Pranzetti D 2010  Black hole entropy from an SU(2)-invariant formulation of Type I isolated horizons  {\it Phys. Rev.} D\textbf{82} 044050

\bibitem{perez-review} Perez A (2017) Black holes in loop quantum gravity {\it Rept. Prog. Phys.} \textbf{80} 126901 

\bibitem{eva1} Ashtekar A and Bojowald M 2005 Black hole evaporation: A New Paradigm {\it 
Class. Quant. Grav.} \textbf{22} 3349--3362 

\bibitem{eva2} Bianchi, E  Christodoulou, M  D'Ambrosio, F  Haggard H and Rovelli C 2018 White holes as remnants: a surprising scenario for the end of a black hole {\it Class. Quant. Grav.} {\bf 35}  225003

\bibitem{eva3} Ashtekar, A  2020  Black hole evaporation: A perspective from loop
quantum gravity, {\it Universe} {\bf 6}  21

\bibitem{eva4} Gambini, R  Olmedo, J and Pullin, R 2020 Spherically symmetric loop quantum gravity: analysis of improved dynamics \texttt{arXiv:2006.01513}

\bibitem{eva5} Kelly, J G Santacruz, R  and Wilson-Ewing E 2020  Black hole collapse and bounce in effective loop quantum gravity \texttt{arXiv:2006.09325} 



\end{thebibliography}
\end{document}